\documentstyle[12pt,draft,epsf]{article}

\newcommand \be  {\begin{equation}}
\newcommand \ee  {\end{equation}}
\newcommand \bea {\begin{eqnarray} \nonumber }
\newcommand \eea {\end{eqnarray}}

\newcommand \de  {\delta}

\newcommand \la  {\lambda}
\newcommand \La  {\Lambda}
\newcommand \s   {\sigma}

\newcommand \cA  {{\cal A}}

\newcommand \refA {($\cA$)}

\begin{document}

\title{Replica Field Theory for Deterministic Models (II): A Non-Random
Spin Glass with Glassy Behavior}
\author{Enzo Marinari$^{(a,b)}$, Giorgio Parisi$^{(c)}$
        and Felix Ritort$^{(a,c)}$\\[0.5em]
  {\small (a): Dipartimento di Fisica and Infn, Universit\`a di Roma
    {\em Tor Vergata}}\\
  {\small \ \  Viale della Ricerca Scientifica, 00133 Roma (Italy)}\\
  {\small (b): NPAC, Syracuse University, Syracuse NY 13210 (USA)}\\
  {\small (c): Dipartimento di Fisica and Infn, Universit\`a di Roma
    {\em La Sapienza}}\\
  {\small \ \  P. A. Moro 2, 00185 Roma (Italy)}\\[0.5em]}
\date{June 8, 1994}
\maketitle

\begin{abstract}

We introduce and study a model which admits a complex landscape
without containing quenched disorder. Continuing our previous
investigation we introduce a disordered model which allows us to
reconstruct all the main features of the original phase diagram,
including a low $T$ spin glass phase and a complex dynamical behavior.

\end{abstract}

\vfill

\begin{flushright}
  {\bf  cond-mat/9406074}\\
  {\bf ROM2F/94/016}\\
  {\bf Roma-La Sapienza 1027}
\end{flushright}

\newpage

\section{Introduction\protect\label{S_INT}}

In a recent companion paper \cite{MAPARI1} (which in the following we will
quote as \refA) we have started (at the same time than Jean Philippe
Bouchaud and Marc Mezard in \cite{BOUMEZ}) a study of the role of replica
field theory when applied to the study of systems which {\em do not} contain
quenched disorder (for further connected work which helps clarifying this
issue see \cite{MIGLIO,MIGRIT}).

The immediate starting point which prompted our investigation \refA\ was a
model of binary sequences with low autocorrelation, as originally discussed
from Golay and Bernasconi \cite{GOLAY,BERNAS}. The model was for us a
prototype of a system which does not contain quenched random disorder, but has
an interesting spin-glass like low $T$ structure (for general discussions
about disordered systems, see \cite{MEPAVI,PARISI,BINYOU}). We have shown that
replica theory allows to gather information about the full phase diagram of
the theory, excluding only the very low $T$ behavior, which is determined by
various factors, including the cardinality of the number of spins of the
system, $N$. We have indeed shown in \refA\  that replica theory can allow a
study of the full deterministic model, and does not have to be limited to an
approximated form.

Apart from such a direct application, we have discussed in \refA\  a more
general valence of such an approach. The ability of investigating
deterministic systems with a complex landscape is an important bonus. We also
stress that we are still lacking a comprehensive description of the glass
state, and that such an approach seems a good candidate to this task.

In the following we will discuss a new class of models without quenched
disorder. They derive quite directly from the ones studied in \refA, by
noticing the peculiar role the Fourier transform is playing (we will discuss
this point in some detail in section \ref{S_GEN}). We will find that these
models behave in a way that appears to be relevant to the description of the
glass state.

We will define the first model (the {\em sine model}) by the Hamiltonian

\be
  \protect\label{E_HS}
  H_S \equiv \sum_{x=1}^{N}
    \{ \sum_{y=1}^{N}[S_x(\sigma_y )] - \sigma_x \}^2\ ,
\ee

\noindent
where

\be
  \protect\label{E_S}
  S_x(\sigma_y) \equiv \frac{1}{\sqrt{N}}
  \sin(\frac{2\pi x y}{N})\  \sigma_y\ ,
\ee

\noindent
and the spin variables $\sigma_x$ take the values $\pm 1$. We define the
analogous {\em cosine model} by the Hamiltonian $H_C$

\be
  \protect\label{E_HC}
  H_C \equiv \sum_{x=1}^{N}
    \{ \sum_{y=1}^{N}[C_x(\sigma_y )] - \sigma_x \}^2\ ,
\ee

\noindent where

\be
  \protect\label{E_C}
  C_x(\sigma_y) \equiv \frac{1}{\sqrt{N}}
  \cos(\frac{2\pi x y}{N})\  \sigma_y\ .
\ee

\noindent
Let us anticipate a discussion of the phase diagram of the model. We
will see that a very important role is played by the case where
$(2N+1)$ is prime (and $N$ is odd for the sine model and it is
even for the cosine
model).  In this case the thermodynamical limit of the partition
function is anomalous. We will show indeed that from the
thermodynamical point of view for prime values of $(2N+1)$ our models
undergo a first order transition at temperature $T_C$. We find such
{\em crystallization} transition only in the case of prime $(2N+1)$. At
$T_C$ the system goes from a disordered state to an highly ordered
one. The specific heat in the low temperature crystalline state is
extremely small.

The system however has a metastable phase whose internal energy is
regular at $T_C$. When we start from high $T$ with a local Monte Carlo
dynamics, and we decrease $T$ with some kind of annealing
procedure, we pass through $T_C$ without any noticeable change in the
thermodynamical quantities.

At a lower temperature $T_G$, within the metastable phase, there is a
transition to a {\em glassy} phase (a second order phase transition). This
transition exists for generic values of $N$. In the  glassy phase the system
may exist in many different equilibrium metastable states. Here there are many
states which survive with finite probability in the infinite volume limit (in
other words replica symmetry is broken). In this phase the system freezes and
thermodynamic fluctuations (for instance of the energy and of the
magnetization) are very small. The behavior of the system at the glass
transition can be  understood in the framework of replica theory.  It is
remarkable that the glass transition temperature $T_G$ is the temperature
where the entropy in the metastable phase becomes nearly equal to the entropy
in the glassy phase (i.e. very close to zero).

We stress again that the crystalline phase exists only for $(2N+1)$ prime, $N$
odd for the sine model and even for the cosine model.  On the other hand the
behavior of the system in the high temperature phase and in the metastable
phase is generic, and does not depend on the cardinality of $(2N+1)$.

In section (\ref{S_GEN}) we will briefly describe the genesis of this model,
after our paper \refA. We will also discuss the low $T$ phase of the low
autocorrelation model, mainly by using number theory. We will again be quite
sketchy, inviting the interested reader to consult \refA\  for a more detailed
discussion. In section (\ref{S_DEF}) we will define a model containing
quenched disorder, which we will eventually dissect by replica theory, and
show to give a fair description of many features of our deterministic models.
We will eventually show that basically the random model and the deterministic
one do coincide, a part for minor details like the non-generic existence of
the crystalline phase in the deterministic models.

In section (\ref{S_REP}) we describe our replica computation.
In section (\ref{S_SAD}) we analyze the saddle point equations. We describe
the replica symmetric and the one step replica broken solution.
In section (\ref{S_MAR}) we discuss the so-called marginality condition.
In section (\ref{S_NUR}) we illustrate our numerical simulations of the
models with quenched random disorder, and in section (\ref{S_NUD}) the
numerical simulations of the deterministic models. In section (\ref{S_MF}) we
discuss the mean field equations for the deterministic models. In section
(\ref{S_CON}) we draw our conclusions. In the final appendix we present the
technical details of a computation concerning the marginal stability.

\section {The Genesis of Our Models\protect\label{S_GEN}}

In order to introduce the models we have defined in the previous section, and
which we will study in the following, let us recall some basic definitions
from \refA, and repeat briefly the reasoning which leads to exhibit the exact
ground state of the model for some particular values of the number of spins.
The reader in need of further details should consult \refA\  and
\cite{PRIMES}.

The low autocorrelation model is based on sequences of length $p$ of spin
variables $\s_x=\pm 1$, with $x=1,p$, and on the Hamiltonian

\be
  \label{E_HLA}
  H = {1 \over p-1} \sum_{k=1}^{p-1} C^2_k\ ,
\ee

\noindent
where the $C_k$ are the correlations at distance $k$, defined as

\be
  C_k \equiv \sum_{j=1}^{p} \s_j \s_{j+k}\ ,
\ee

\noindent
where we are taking periodic boundary conditions (this is, in the terminology
of \refA, the {\em periodic} model), i.e. the indices are always summed modulo
$p$. In this way the indices which address the $\sigma$ variables always
belong, as they should, to the interval $[1,p]$. It is useful to  rewrite the
Hamiltonian as

\be
  \label{E_HLA_FOURIER}
  H = {1 \over p-1}\sum_{k=1}^{p} \Bigl ( |B(k)|^4 -1 \Bigr ) + 1\ ,
\ee

\noindent
where the Fourier transform is defined as

\be
  B(k) \equiv {1 \over \sqrt{p}} \sum_{x=1}^{p}
  e^{ i \frac{2\pi k}{p} x} \s_j\ ,
\ee

\noindent
and $i$ is the imaginary unit. The thermodynamics of the model can be
reconstructed thanks to the partition function at inverse temperature
$\beta \equiv \frac{1}{T}$ in the volume $p$

\be
  Z_p(\beta) \equiv \sum_{\{\sigma\}} e^{-\beta H\{\sigma\}}\ .
\ee

An interesting way to look at the Hamiltonian (\ref{E_HLA}) is to consider it
as a particular form of a fully frustrated $4$-spin interaction. Here
only the $4$ spin terms which are contained in a square of two points
correlation functions appear. This point of view has been useful in \refA\  to
show that replica theory can be a reasonable tool to investigate
deterministic models.

It is remarkable that for prime values of $p$, such that $p=4n+3$, it is
possible to exhibit in an explicit way one ground state of the system. Let us
construct such ground state configuration.
Following Legendre \cite{PRIMES} we set $\s_p=0$ and

\be
  \s_j= j^{{1 \over 2}(p-1)}\  {\rm mod}\  p\ .
\ee

\noindent
In this way  $\s_j$ is $+1$ or $-1$, if $j<p$. Indeed
a theorem by Fermat \cite{PRIMES} tells us that if $j$ is not multiple of $p$,
$j^{(p-1)}=1$, mod $(p)$ and therefore $ j^{{1 \over 2}(p-1)}=\pm 1$.

We will evaluate the energy of this sequence and only at the end
we will impose that $\s_p= \pm 1$ on the last site $p$.  It is well known
that for this sequence all the correlations $C_k$ are equal to
$-1$ \cite{PRIMES}. It is also remarkable (and the crux of this paper)
that on such a sequence the Fourier transform is given by

\be
  \label{E_GAUSS}
  B(k) = G(p)\, \s_k \ ,
\ee

\noindent
where, according to Gauss \cite{PRIMES}, $G(p)=1$ for $p=4n+1$ and
$G(p)=-i$, for $p=4n+3$. This Gauss theorem makes easy to verify that the
configurations we have exhibited have energy $1$ (the lowest possible energy
for odd values of $p$). Now  we  change the last spin to $\pm 1$. It is easy
to verify that after doing that the energy of configurations with $p$ of the
form $4n+3$ stays unchanged to $1$, while for $p=4n+1$ the energy grows to
$5$. It is clear now that for $p$ prime of the form $4n+3$ we have exhibited
a true ground state of the low autocorrelation model.

By using Gauss theorem about Fourier transforms of Legendre sequences we are
able now to define a simple model with $2$-spin interaction which has the same
ground state of the $4$-spin interaction low autocorrelation model. We are
ignoring here the presence of the spin with value zero. The new Hamiltonian
has the form

\be
  \protect\label{E_HQUA}
  H=\sum_{x} |G(p)\, \s_x- B(x)|^2\ .
\ee

We can further simplify the model by noticing that the sequence of the
$\s$ in the ground state we have written is symmetric or antisymmetric around
the point $\frac{p-1}{2}$, depending on the value of $G(p)$. That allows us
to define two new models with half the number of degrees of freedom which
continue to admit (for selected $p$ values) the ground state we have written.
Such two models are exactly the sine and the cosine model we have define in
our introduction.

Hopefully we have given clarifying hints about the nature of our two models.
Now we can proceed to study them.

\section{The Disordered Model\protect\label{S_DEF}}

It is natural to introduce at this point a model which contains quenched
disorder. The companion paper \refA\ justifies in detail this approach. By
studying a suitable disordered model we try to understand how general is a
very specific $2$-spin interaction like for example the sine
one (\ref{E_HS}). We will find they have indeed much in common, and that the
random model allows to reconstruct exactly the most part of the phase
diagram. As before we define the Hamiltonian (here $O$ stands for orthogonal)

\be
  \protect\label{E_HO}
  H_O \equiv \sum_{x,y=1}^{N} O_{x,y}\sigma_x\sigma_x \ ,
\ee

\noindent
where now $O_{x,y}$ is a generic orthogonal symmetric matrix.
The same behavior of the
deterministic model will be obtained by using a rescaled Hamiltonian

\be
  \protect\label{E_HTO}
  \tilde{H}_O \equiv 2N- 2H_O\ .
\ee

\noindent
The form we have just written is important since also in the case of the
original sine and cosine models the Hamiltonians defined after
eq.(\ref{E_HQUA}) can be written in the form  $ 2N-2\sum_{i,k} M_{i,k} \s_i
\s_k$, by neglecting terms which are irrelevant in the $N \to \infty$
thermodynamic limit.

The first element for the comparison of the two class of models, the sine and
cosine versus the random one, can be obtained from noticing general features
of the high temperature expansions of the models. For both class of models the
couplings\footnote {This is not true for all soluble spin glass models. In the
dilute models the average coordination number $z$ remains finite and the
couplings may be quantity of order $1$, with a probability of order
$\frac{z}{N}$.} are of order $N^{-\frac{1}{2}}$. The diagrams which contribute
to the infinite volume limit have the same topology for the two classes of
models, and they only depend on quantities like the trace of the couplings
to positive powers, which have been built to be equal in the two classes of
models.

The reasoning of the former paragraph proves that sine and cosine models
defined from  (\ref{E_HS}) and (\ref{E_HC})  and the model with quenched
disorder defined from (\ref{E_HTO}) have the same high temperature expansion.
On the other side we have exhibited a ground state of the deterministic system
which exists for prime values of $(2N+1)$. Such construction obviously does
not apply to the disordered models. This implies that the static properties
of the two class of models (for prime values of $(2N+1)$) cannot
coincide all the way down to $T=0$. There is a crystallization transition
only in the deterministic models, thanks to very peculiar cardinality
properties of $N$.

We will give evidence that the random and the deterministic model do
coincide at all temperatures in the metastable phase. This is the case
for generic values of $N$, since as we already stressed the cardinality of
$2N+1$ is irrelevant for the behavior of the deterministic model in the
metastable phase. A similar pattern could hold for the low
autocorrelation model, but in the present case of the $2$-spin interaction
the analysis is far simpler, and we are able to carry it through all the way.

\section{The Replica Approach\protect\label{S_REP}}

By using replica theory techniques \cite{MEPAVI,PARISI} we will solve now the
model with quenched disorder defined by the Hamiltonian (\ref{E_HO}). As usual
we define the free energy of $n$ replicas as

\be
  \protect\label{E_FN}
  f^{(n)}(\beta) \equiv
  \lim_{N \to \infty} \Bigl (
  -{1 \over \beta N} \frac{\overline {Z_O(\beta)^n}-1}{n} \Bigr ) \ ,
\ee

\noindent
where with the bar we denote the average over the quenched disorder and

\be
  Z_O^n \equiv \sum_{\{\s^a\}} \exp \{-\beta \sum_{a=1}^n H_O^a\}\ .
\ee

\noindent
We have to average over the quenched disorder. To this
end we have to compute

\be
  \protect\label{E_INTOZERO}
  \overline{Z_O^n}=
  \int dO \exp\{\sum_{k,j=1}^{N}\beta\  \Omega_{k,j} O_{k,j}\}\ ,
\ee

\noindent where the integral runs over orthogonal symmetric matrices, and

\be
  \protect\label{E_OMEGA}
  \Omega_{k,j} \equiv \sum_{a=1}^{n} \s^a_k \s^a_j \ .
\ee

\noindent
We will show now that we can solve a more general problem considering a
symmetric coupling matrix with some quite general preassigned eigenvalue
distribution. We will derive such more general form.  We will eventually
obtain the relevant result specializing this general form to orthogonal
symmetric matrices.

A generic real symmetric matrix $O$ can be decomposed as\footnote{We like to
stress with $^*$ the operation of hermitian conjugation, which for real
matrices coincide with transposition.}

\be
  O = V D V^* \ ,
\ee

\noindent
where $D$ is a diagonal matrix which controls the spectrum of $O$, and
$V$ is the orthogonal matrix which diagonalizes $O$. By using this
decomposition we have to compute

\be
  \overline{Z_O^n}
  = \int dV\  \exp\{ \mbox{\rm Tr} (\beta V \Omega_{k,j} V^* D) \}\ ,
\ee

\noindent
where $D$ is a diagonal matrix, $dV$ is the Haar invariant measure over the
orthogonal group, and the matrix $\Omega$ is defined in (\ref{E_OMEGA}). We
can use the results derived in \cite{ITZZUB} for unitary matrices and adapt
them to the orthogonal case. So, let us assume for a while that we are
integrating over unitary matrices $V$. Using the fact that $\Omega$ {\em has
finite rank} we find that

\be
  \int dV\  \exp\{ \mbox{\rm Tr} (\beta V \Omega_{k,j} V^* D) \}
  =\exp \{N \mbox{\rm Tr}  G_D({\beta \Omega \over N})\}\ .
\ee

\noindent
The value of $G$ is given in \cite{ITZZUB} (when, as we already said, the
integral is over the unitary matrices). Following \cite{ITZZUB} let us define
the generating function for the traces of $D$ as

\be
  \Phi_D(j) \equiv \frac{1}{N} \sum_{k=0}^\infty j^k \mbox{\rm Tr} D^k\ ,
\ee

\noindent in the case where $d\equiv\mbox{\rm Tr} D = 0$. If $d\ne 0$ we
define the generating functional as

\be
  \Phi_D(j) \equiv \frac{1}{N} \sum_{k=0}^\infty j^k \mbox{\rm Tr} (D-d)^k\ ,
\ee

\noindent that allows a straightforward generalization of the computation, by
only adding an additional contribution to the free energy. We define the
function $z_D$ as

\be
  \protect\label{E_ZD}
  z_D(j) \equiv j \Phi_D(j)\ ,
\ee

\noindent and finally we define the function $\psi_D(z)$ by

\be
  \psi_D(z) \equiv \Phi(j_D(z))\ ,
\ee

\noindent where $j_D(z)$ is obtained by inverting (\ref{E_ZD}). All said,
\cite{ITZZUB} tells us that $G$ is given by

\be
  \protect\label{E_PSIG}
  G_{D}(z) = \int_0^1 dt\ \frac{\psi_{D}(zt)-1}{t}\ .
\ee

\noindent
In the orthogonal symmetric case $O^2=1$ and the eigenvalues of $D$  can take
the values $\pm 1$. As far as our problem is concerned we are interested in the
case where half of the eigenvalues take the value $+1$ and half the value
$-1$. We will discuss here a more general case, where  a fraction $\nu$ of the
eigenvalues is $+1$ and a fraction $1-\nu$ is $-1$.

It is interesting to notice that the ground state of the model has a simple
geometrical significance.  Let us consider our series of $N$ spins
$\sigma$, and look at it as one of the vertices of the unit hypercube in $N$
dimensions. Let us imagine such an hypercube as embedded in $\Re^N$. Now we
extract a random linear subspace $F$ of dimension $\nu N$, which includes the
origin. For example if we have $N=2$ spins the configuration will seat on one
of the four corners of a $2$ dimensional square, and for $\nu=\frac{1}{2}$ we
would pick a random line passing through the origin.
If $P$ is the projector of $F$ the matrix $O$ is given by

\be
  O = 2 P -1\ .
\ee

\noindent
We define the norm of
the projection of a  spin configuration $\{\s\}$ over the subspace $F$ by

\be
  P_\s= |P \s|\ ,
\ee

\noindent
and the norm of the projection over the complementary subspace $F^{\perp}$

\be
  D_\s = |(1 - P) \s|\ .
\ee

\noindent
$D_\s$ can be interpreted as the distance of the configuration $\s$ from the
subspace $F$. The relation $P_\s^2+D_\s^2=1$ holds.
The Hamiltonian (\ref{E_HTO}) can be written now as $4 D_\s^2$. The ground
state energy is given by the minimum distance $D_m$ of one of the $2^N$
configuration from the random subspace.  This problem is well studied in the
case $\nu N =1$, i.e. in the limit $\nu \to 0$, mainly for its applications
to perceptrons \cite{PERCEP}, but it has not been discussed in the most
general case.

For $\nu=1/2$ by inverting the second relation after some algebra we find
(we omit the suffix $\nu=\frac{1}{2}$ for $G$ and $\psi$)

\be
  G(z)=\int_0^1 dt\frac{\sqrt{1+4z^2t^2}-1}{2z}\ ,
\ee

\noindent
which gives

\be
  G'(z)=\frac{\psi (z)-1}{z}\ .
\ee

\noindent
After integrating the last relation with the condition $G(0)=0$ we find

\be
  \protect\label{E_PRIMA}
  G(z)=\frac{1}{2}\log(\sqrt{1+4z^2}-1)-\frac{1}{2}\log(2z^2)+
  \frac{1}{2}\sqrt{1+4z^2}-\frac{1}{2}\ ,
\ee

\noindent
where the constant term has be chosen such that  $G(0)=0$.

We have already said that we have obtained this $G$ for $V$ unitary. It is
easy to argue that when we integrate over orthogonal matrices the only
difference is that $G(\beta z)$ gets substituted from
$\frac{1}{2}G(2\beta z)$. That can be seen for example by noticing that the
function $G$ has to be the same in the two cases (since the same diagrams
contribute) and at first order in $\beta$ orthogonal and unitary matrices have
to give the same results. So the only allowed renormalization will be of the
kind $G(z) \to \alpha G(\frac{z}{\alpha})$. The counting of the eigenvalues
leads to the conclusion $\alpha=\frac{1}{2}$.

Using the fact that for integer positive $k$

\be
  \mbox{\rm Tr} \Bigl ( \frac{\beta\Omega}{N}\Bigr )^k =
  \mbox{\rm Tr} \Bigl ( \beta\Sigma          \Bigr )^k \ ,
\ee

\noindent where the matrix $\Sigma$ is defined as

\be
  \Sigma_{a,b} \equiv \sum_{k=1}^{N} \s^a_k \s^b_k \ ,
\ee

\noindent it follows immediately that

\be
  \mbox{\rm Tr} G(\frac{\beta\Omega}{N}) =
  \mbox{\rm Tr} G(\beta\Sigma)\ .
\ee

\noindent
To continue our computation we insert a $\delta$-function, and introduce the
Lagrange multipliers $\Lambda$ with the representation

\be
  \prod_{a,b=1}^{n} \de( \sum_{j=1,N}\s^a_j \s^b_j -N Q_{a,b}) \simeq
  \int \prod_{a,b=1}^{n} d \La_{a,b}\  \exp\{i\sum _{a,b}
  \La_{a,b}(\sum_{j=1,N}\s^a_j \s^b_j -N Q_{a,b})\}\ .
\ee

\noindent
After a little more algebra (very similar to the one developed in \refA) we
find that

\be
  \overline{Z^n}=\int dQ\,d\La\,\exp(-N A[Q,\La])\ .
\ee

\noindent
In the large $N$ limit the free energy is obtained by finding the saddle
point value of $A[Q,\La]$, which has the form

\be
  \protect\label{E_ASP}
  A[Q,\La] = - \frac{1}{2} \mbox{\rm Tr} G(2 \beta Q)
  + \mbox{\rm Tr} (\La Q) - F(\La)\ ,
\ee

\noindent where $G$ has been already defined, and

\be
  \protect\label{E_FLA}
  F(\La) \equiv \ln \sum_{\s^a} \exp\{ \sum _{a,b}\La_{a,b} \s^a \s^b\}\ .
\ee

\noindent
We will need to study eq. (\ref{E_ASP}) to discuss the
solutions of the model.

\section {Saddle \ Point\ \  Equations and Replica Symmetry
Breaking\protect\label{S_SAD}}

In the previous section we have found the saddle point equations which
allow to solve the model with quenched disorder defined in (\ref{E_HO}). Let
us recall that
the free energy (multiplied times $n \beta$) in terms of the matrices
$Q$ and $\La$ is

\be
  \protect\label{E_FO}
  \beta f_O = \lim_{n\to 0} \frac{A[Q_{SP},\La_{SP}]}{n}\ ,
\ee

\noindent
where $A$ is defined in (\ref{E_ASP}), and $Q_{SP}$ and $\La_{SP}$ are
evaluated at the saddle point of $A$.

The free energy (\ref{E_HTO}) of the model with quenched random disorder
(which has the same high-temperature expansion than the deterministic one
(\ref{E_HS},\ref{E_HC})) is given by

\be
  \beta f = 2\beta - 2\beta f_O(2\beta)\ .
\ee

\noindent
Let us start by considering the annealed case,  $n=1$. Here the matrix
$\Sigma$ is set equal to $1$. The action does not depend on $\La$, and we
find for the free energy density and the internal energy

\bea
  \protect\label{E_EQRS}
  f &=& 2-\frac{1}{2\beta}G(4\beta)-\frac{1}{\beta}\log(2)\ ,\\
  e &=& 2(1-G'(4\beta))= 2-\frac{\sqrt{1+64\beta^2}-1}{4\beta}\ .
\eea

\noindent

\begin{figure}
  \epsffile[60 206 565 690]{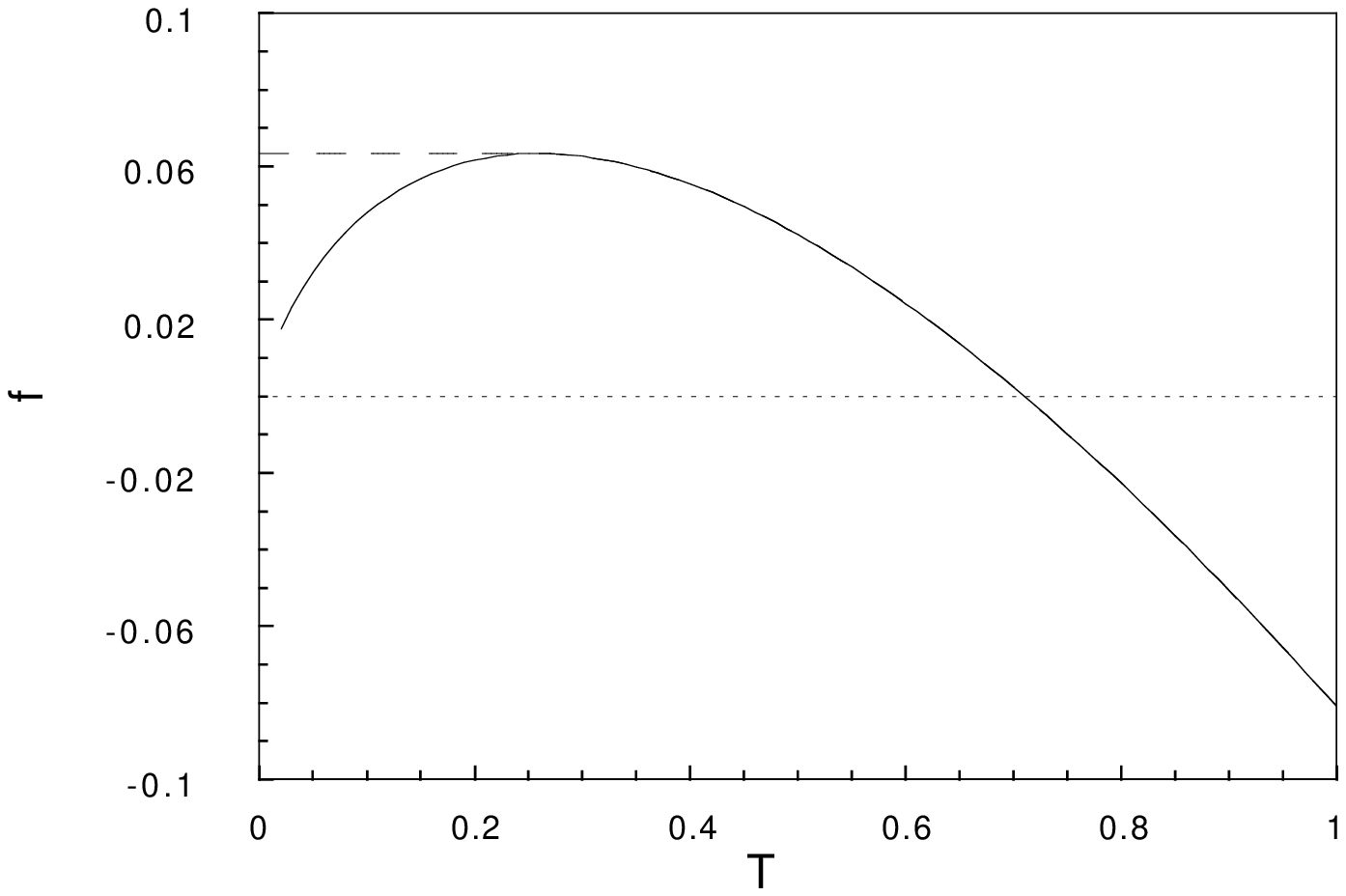}
  \caption[a]{\protect\label{F_RSFREE}
  Free energy of the model with quenched random disorder versus $T$. The
  continuous line is the replica symmetric solution, the dashed line is
  the one step replica broken solution. With the dotted line we only
  indicate the zero of the free energy.
  The free energy vanishes at $T\sim 0.71$.}
\end{figure}

We plot the replica symmetric free energy found in eq.
(\ref{E_EQRS}) in fig. (\ref{F_RSFREE}) (together with the one step
replica broken result we will compute in the following).

\begin{figure}
  \epsffile[60 206 565 690]{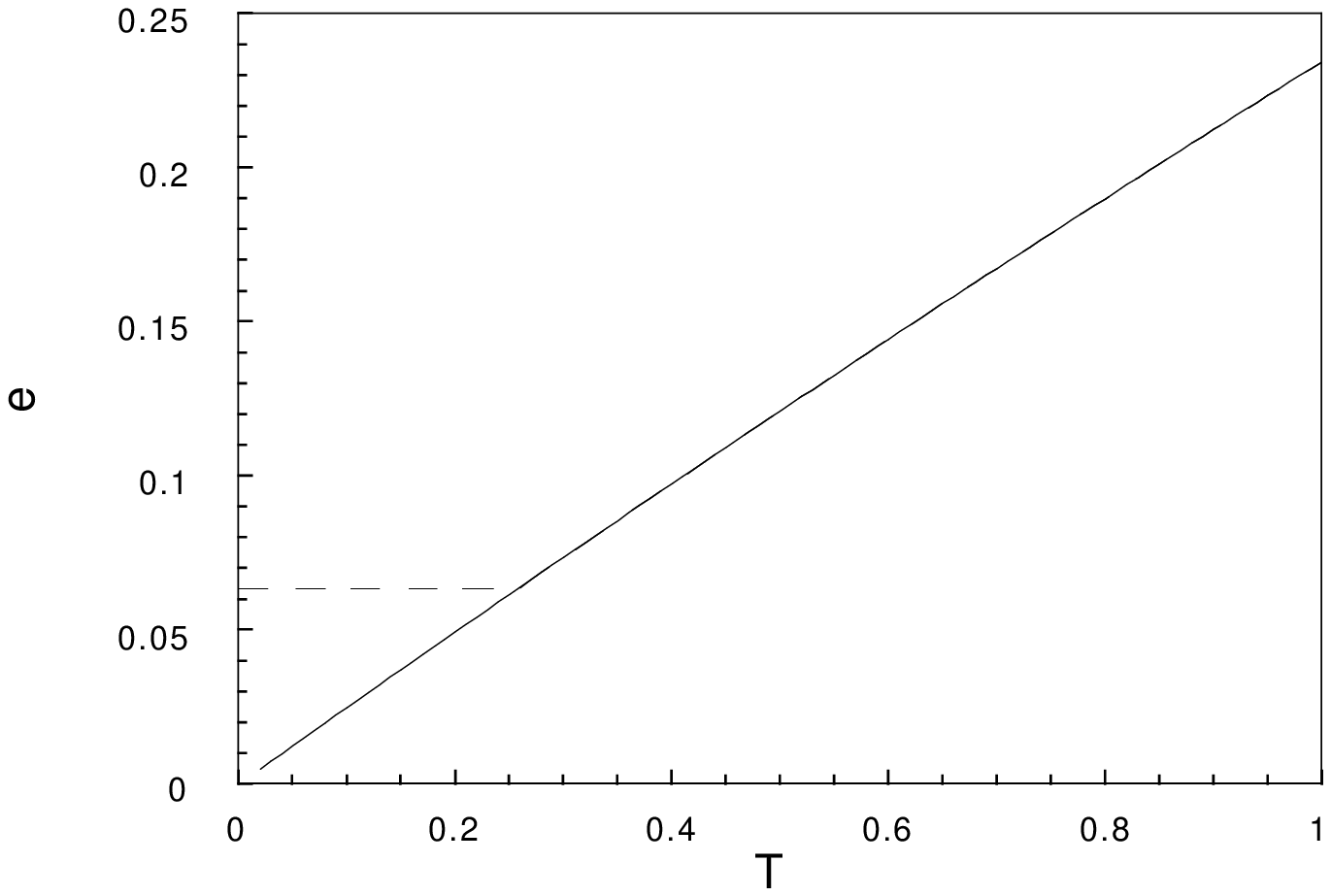}
  \caption[a]{\protect\label{F_RSENER}
  As in fig. (\ref{F_RSFREE}), but for the internal energy of the system.}
\end{figure}

\begin{figure}
  \epsffile[60 206 565 690]{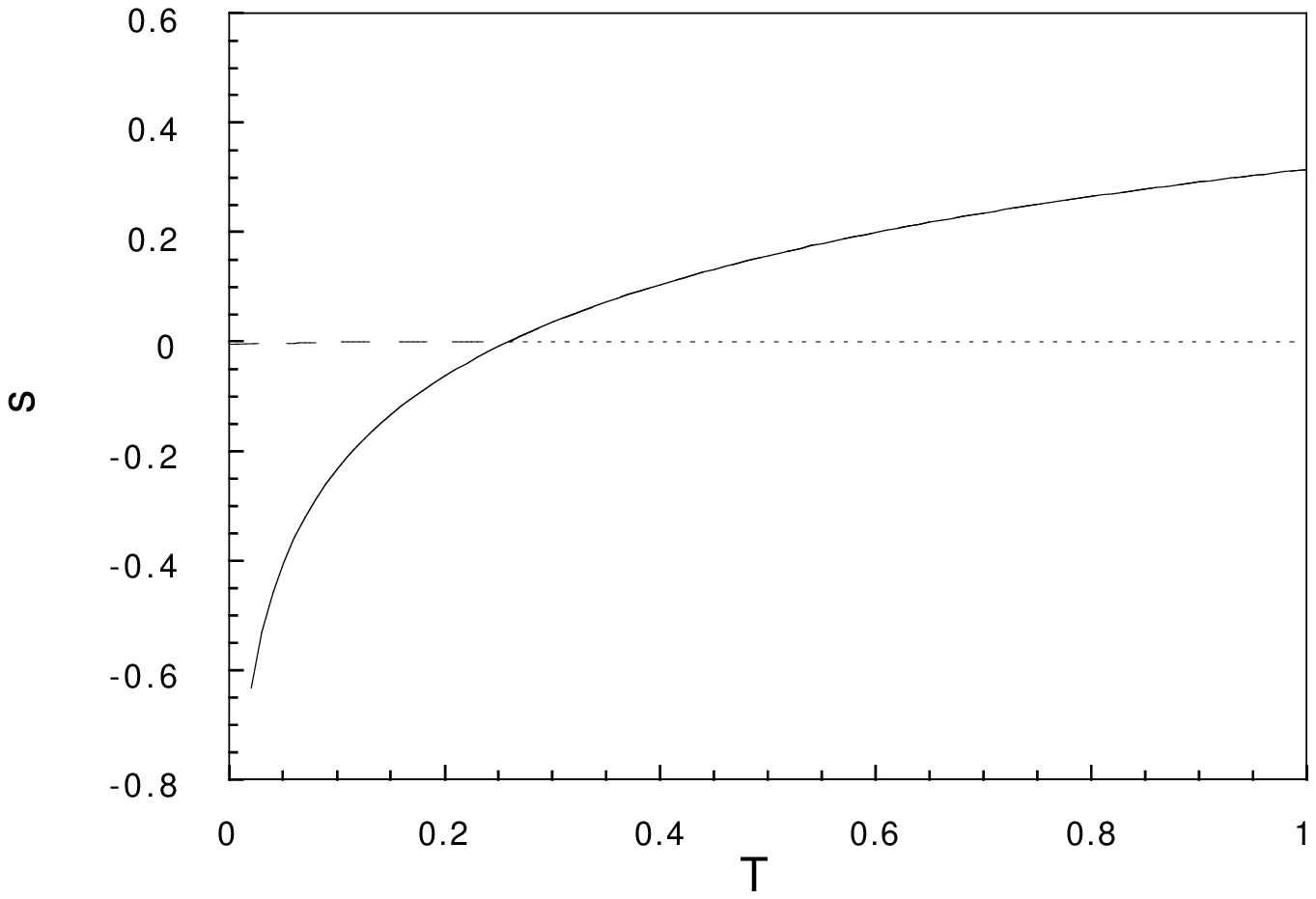}
  \caption[a]{\protect\label{F_RSENTR}
  As in fig. (\ref{F_RSFREE}), but for the entropy of the system. Here
  again the dotted line is solely meant to indicate the zero.
  The entropy of the one step solution is
  very small in the low $T$ phase. }
\end{figure}

In fig. (\ref{F_RSENER}) we plot the internal energy and in fig.
(\ref{F_RSENTR}) the entropy of the system.

In the high-temperature region the quenched and the
annealed solutions coincide as usual for long range models.

The replica symmetric solution is stable at all temperatures. But
since for $T \le 0.26$ it gives a negative entropy (see
fig. (\ref{F_RSENTR}) it cannot be correct down to $T=0$. We expect
replica symmetry to break above (but very close) to $T=0.26$. Here the
system enters a glassy phase very similar to that of the random energy
model \cite{REM} and of the $p$-spin systems (see for example
\cite{CRCUKU,KIRTHI}).

We can compute the one step replica broken solution.  We parameterize
the matrices $Q$ and $\La$ in the usual way. In presence of an uniform
magnetic field the matrix elements $Q_{ab}$ take the value $q$ if $a$
and $b$ belong to the same sub-block of size $m$, while they take the
value $q_0$ if they do belong to different sub-blocks.  We parameterize
the matrix $\La_{ab}$ with blocks of the same size $m$, and we set its
elements equal to $\la$ or $\la_0$ with the same procedure we used for
$Q$. We consider here the simpler case of zero magnetic field,
where the parameters $q_0$ and $\la_0$ are zero and we set

\be
Q_{a,b}=q~~(a\ne b), \ \ \ \ \Lambda_{a,b}=\la~~ (a\ne b)
\label{break}
\ee

\noindent
inside the blocks of size $m$ ($Q_{aa}=1;\La_{aa}=0$).
After some algebra we obtain

\bea
  \label{E_1SBFREE}
  \beta f=2\beta -\frac{1}{2m}\,[(m-1)G(4\beta (1-q))+G(4\beta (mq+1-q))]\\
  +\la q(m-1)-\log(2)\,+\la-
  \frac{1}{m}\log \int_{-\infty}^{\infty}
  \,\frac{dx}{\sqrt{2\pi}}e^{-\frac{x^2}{2}}ch^m(\sqrt{2\la}\,x)\ .
\eea

\noindent
The stationary equation for $q$ tells us that

\be
  \label{E_1SBLAMB}
  \la=\frac{2\beta}{m}\,[G'(4\beta (mq+1-q))-G'(4\beta (1-q))]\ .
\ee

\noindent
We can use this relation to eliminate $\lambda$ from
(\ref{E_1SBFREE}). We are left with a a function of $q$ and $m$, and
we have to find a stationary point. This expression cannot be solved
in close form. We have plotted the numerical solution with dashed lines in
figures (\ref{F_RSFREE}), (\ref{F_RSENER}), (\ref{F_RSENTR}).

At $T_{RSB}\sim 0.26$ there is a phase transition to a phase with
broken replica symmetry. At the transition point $T_{RSB}$ the value
of the entropy is finite but very small ($\sim 0.0004$), the value of
$q$ jumps discontinuously to a value very close to $1$ ($\sim 0.9998$),
and $\la$ is large but finite ($\sim 10$) (in the Random Energy Model
at the transition point $q=1$ and $\la=\infty$ \cite{GROMEZ}).  Below
$T_{RSB}$ the parameter $m$ is very approximately proportional to $T$,
$m=1$ at $T_{RSB}$.  This is the typical scenario for a large class of
models where the order parameter jumps discontinuously at the
transition.

We have not studied in detail the stability properties of the replica
broken solution. It is possible that the one step solution is stable
down to a very low temperature, and that for lower values of $T$ a
continuous symmetry breaking is needed to describe the system. This is
what happens for the $p$-spin model \cite{GARDNER}.  As we will
discuss in the next sections this second transition would probably
have no relevance from the physical point of view, since the system is
not able to explore the lowest free energy configurations. We will see
that in an usual annealing process (i.e., a slow temperature cooling
starting from a high temperature) the system has a transition at a
temperature $T_G$ well above the temperature $T_{RSB}$ where replica
symmetry breaks down. We will name the transition at $T_G$ the {\em
glass transition}. This transition is dynamical in nature and
corresponds to the presence of a very large number of metastable
states.  At $T_G$ the system remains trapped in a metastable state,
and thermal fluctuations are very small.

\section{The Marginality Condition\protect\label{S_MAR}}

In the framework of mean-field theory it has been suggested that the
solution to the Sompolinsky-Zippelius dynamical equations
\cite{SOMZIP} undergoes a phase transition at a temperature $T_G$.
Below that temperature the time-homogeneity hypothesis and the
standard fluctuation-dissipation theorem are not valid. In the SK
model the temperature $T_{G}$ coincides with the transition point
derived from the static approach, where the replica symmetric solution
becomes unstable. It has also been suggested that this temperature
coincides with the temperature $T_{MC}$ where a one step replica
broken solution to the mean field equations exist such that the size
of the replica matrix sub-block $m$ is fixed by the condition that the
replicon eigenvalue vanishes. This has been called the {\em
marginality condition} \cite{SOKODO}.

More recently several authors have investigated the $p$-spin spherical
spin glass model \cite{CRCUKU}. In this case it is possible to write
closed expressions for the correlation and response functions in the
off-equilibrium regime. It has been noted \cite{KIRTHI} that the
dynamical equations undergo a glass transition at a temperature $T_G$
where the relaxational dynamics slows down and aging effects start to
appear. The temperature $T_G$ is larger than the transition point
where replica symmetry breaks down, as predicted by the static
approach.  This is a consequence of the stability of the replica
symmetric solution and corresponds to the fact that at the transition
point the spin-glass order parameter $q(x)$ is discontinuous. In this
model $T_G$ coincides with $T_{MC}$.

The models we are describing in this work (the model with quenched
random disorder as well as the deterministic one) are good candidates
for a test of the marginality condition principle. The main reason is
that at the transition point the order parameter $q$ jumps
discontinuously to a value extremely close to $1$. The system
essentially freezes and the difference between the static transition
temperature $T_{RSB}$ and the dynamical transition temperature value
$T_G$ is large.  In the following sections we will use numerical
simulations to show that, for reasons not completely clear to us, the
principle seems to work well.

Now we want to derive the value of $T_{MC}$ in our particular case.
We start from eq.(\ref{E_ASP}) and we compute the Hessian matrix in the
$\La,Q$ space. The interested reader can find the technical details
in the appendix.  The marginality condition gives

\be
  \label{E_MARG}
  16\beta^2\,G''(4\beta(1-q))
  \langle\cosh(\sqrt{2\lambda}x)^{-4}\rangle=1\ ,
\ee

\noindent where the expected value is defined by

\be
  \langle A(x)\rangle =\frac{\int dx \frac{e^{-x^2}}{\sqrt{2\pi}}
  \cosh^m(\sqrt{2\la}x)A(x)} {\int dx \frac{e^{-x^2}}{\sqrt{2\pi}}
  \cosh^m(\sqrt{2\la}x)}\ .
\ee

\noindent
We can find the dynamical transition point by maximizing the free
energy under the marginality condition.

\begin{figure}
  \epsffile[60 206 565 690]{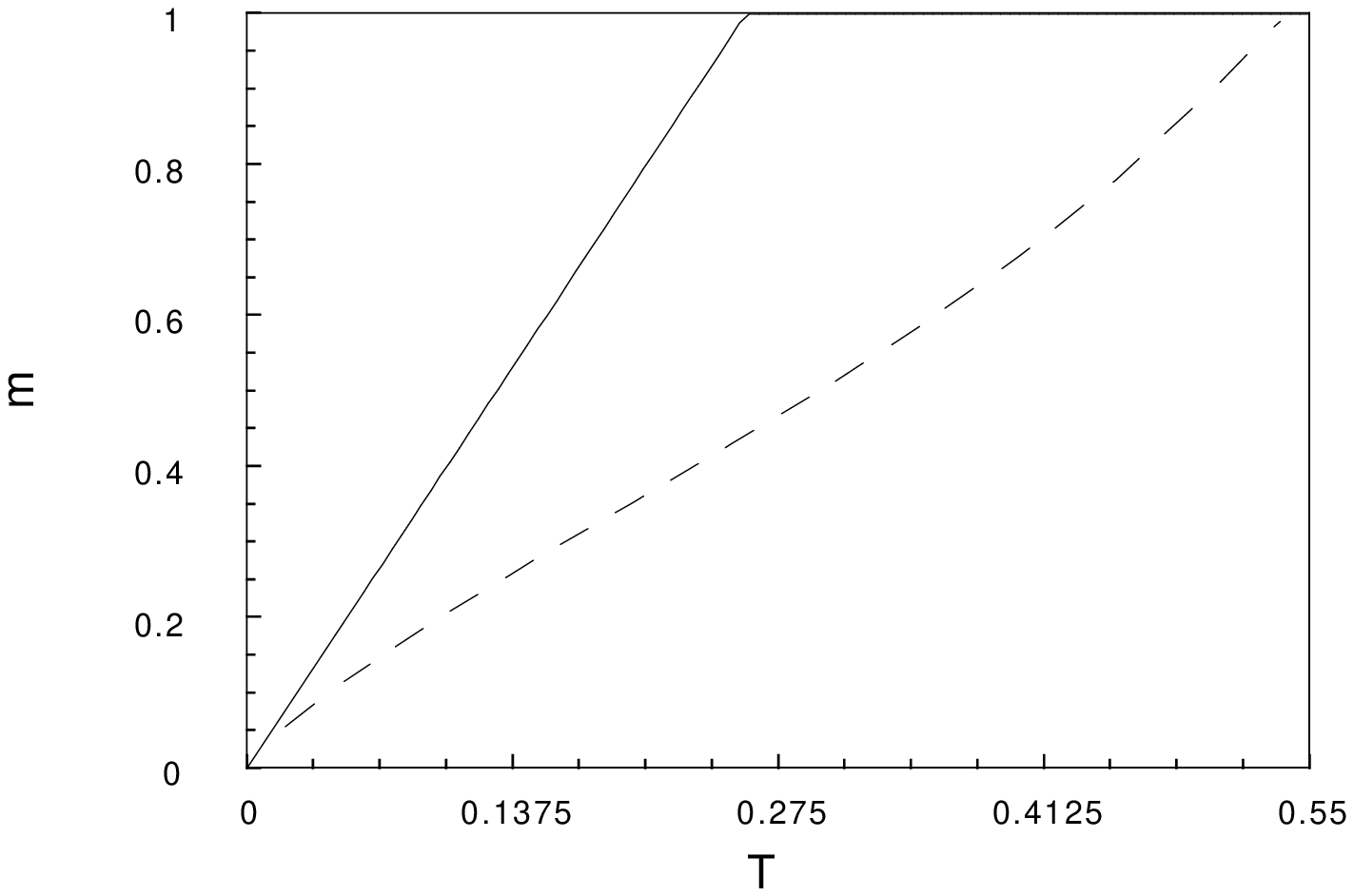}
  \caption[a]{\protect\label{F_MMAR}
  Replica matrix sub-block size $m$ as a function of $T$. The continuous
  curve is for the static value, the dashed curve is
  for the solution satisfying the marginality condition.}
\end{figure}

Maximizing the free energy (\ref{E_1SBFREE}) as a function of $q$ for
$m$ fixed, under condition (\ref{E_1SBLAMB}), we find that there are
values of $m\le 1$ such that eq. (\ref{E_MARG}) is satisfied as soon
as $T\le T_{MC} \simeq 0.535 \pm 0.005$.  This transition temperature
is two times larger than $T_{RSB}$.  We also get $m$ and $q$ as a
function of the temperature.  At $T_{MC}$ $q$ jumps discontinuously to
a value$\simeq 0.962$. This value is smaller than the value we have
found for the static solution. A priori we cannot expect the free
energy derived using the marginality condition principle to a
reasonable quantity, i.e. to satisfy the main inequalities of the
thermodynamics. This is because we are in the wrong branch of the
solutions of the replica equations, and we have not chosen $m$
following a variational principle.  For example the relation
$u=\frac{\partial (\beta f)}{\partial \beta}$ is not satisfied for the
marginality condition free energy. Also the value of the breakpoint
parameter $m$ (which we plot in fig. (\ref{F_MMAR}) together with the
value from the static result) is not proportional to $T$ at low
temperatures.

It is not possible to describe the behavior of the system in the
glassy phase without solving the full off-equilibrium equations,
except for the value of the glassy temperature.
As we have already discussed a complete analysis should not be
confined to the case of one step replica symmetry breaking step.  It
would be very interesting to analyze the full low $T$ behavior for a
larger number of breaking steps, and eventually for a continuous
breaking pattern.  In the following sections we will present a
numerical study of the model with quenched disorder and of the
deterministic model. We will see that in both cases the system undergoes a
dynamical transition at $T_G$, and that $T_G$ is very close to the
value $T_{MC}$ we have computed here.

\section{Numerical Simulations of the Disordered
Model\protect\label{S_NUR}}

The model with quenched disorder is based on symmetric orthogonal
interaction matrices. In order to produce the interaction matrices
needed in our simulations we started by generating a
symmetric matrix with random elements with a Gaussian distribution.
Starting from such a matrix we have obtained a symmetric orthogonal matrix
by using the Graham-Schmidt orthogonalization procedure.

Such a model has an infinite range interaction, and Monte Carlo
simulations are quite time consuming (but much less time consuming
than for example $p$-spin models with $p>2$). With limited computer
time (on a reasonable workstation time allocation) we have been able
to obtain reliable results for samples with a volume up to a few
hundred spins.

\begin{figure}
  \epsffile[60 206 565 690]{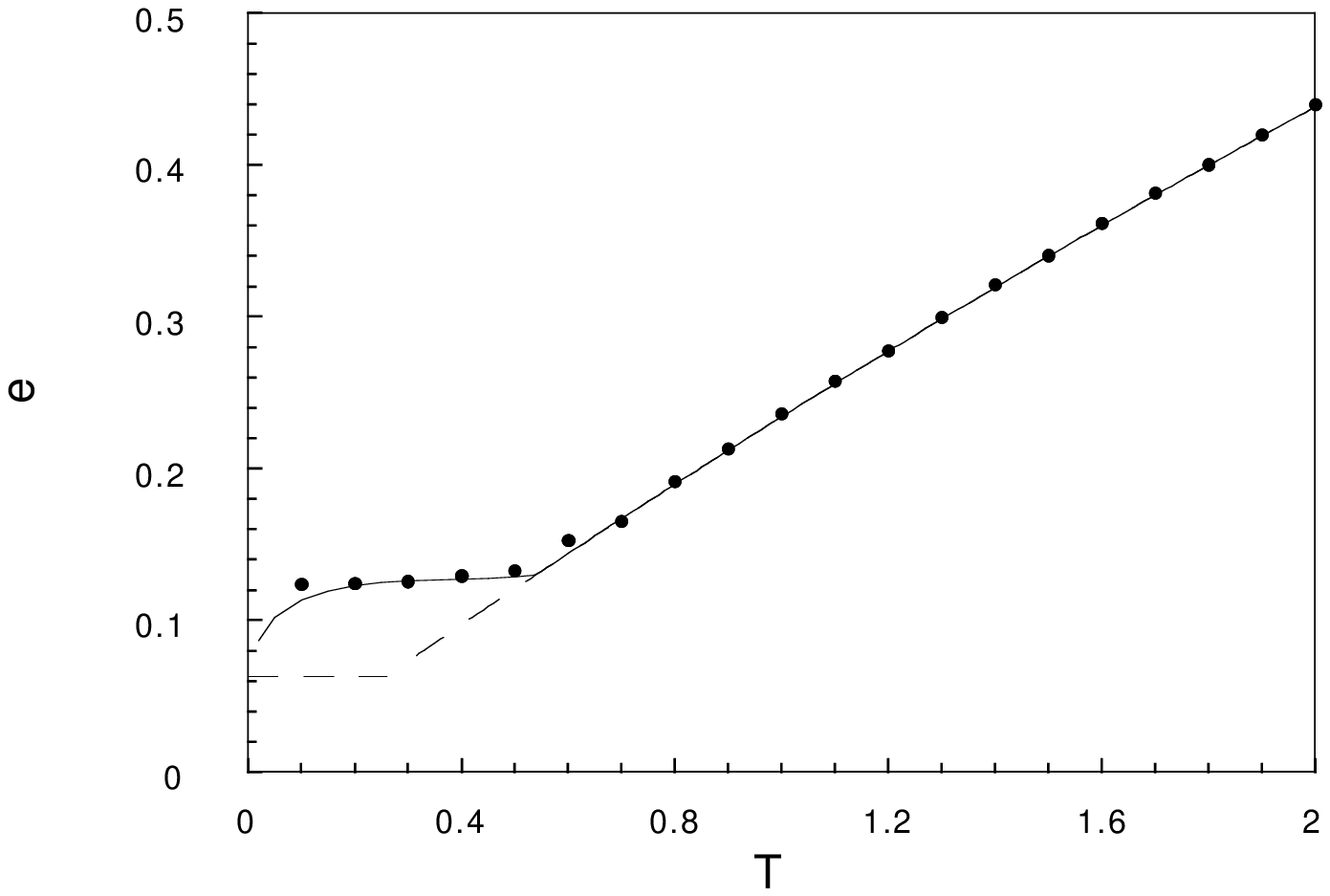}
  \caption[a]{\protect\label{F_ENRANU}
Energy of the model based on random orthogonal matrices
versus $T$ for $N=186$. The
dashed line is the static one step replica broken solution. The
continuous line is the prediction of (\ref{E_MARG}).
}
\end{figure}

\begin{figure}
  \epsffile[60 206 565 690]{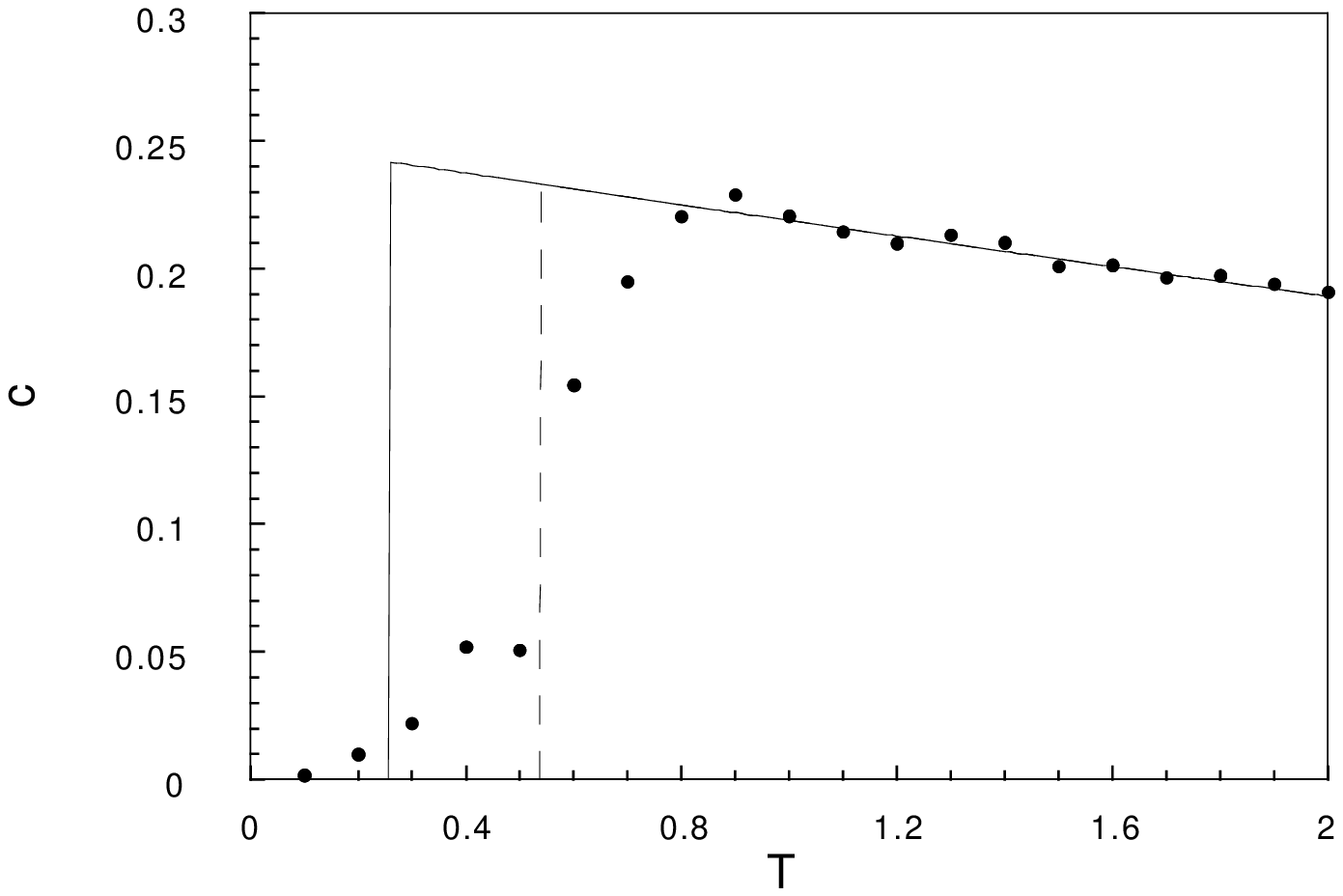}
  \caption[a]{\protect\label{F_CVRANU}
  As in figure (\ref{F_ENRANU}), but for the specific heat. Here the
  continuous line is  the static one step replica broken solution,
  while the dashed line is the prediction of (\ref{E_MARG}) (inverting
  the notation of the former figure).
}
\end{figure}

In figure (\ref{F_ENRANU}) we show our estimate for the internal
energy on one disorder sample, for $N=186$. In figure (\ref{F_CVRANU})
we show the specific heat. We have started the run from high $T$ and
we have been decreasing the temperature at steps of $.1$.

We have tested that sample to sample fluctuations
and finite-size corrections in the internal energy and heat capacity
are negligible.

Our numerical results fit well the theoretical predictions for
temperatures larger than $T_G\sim 0.5$. At $T_G$ the system
freezes.  The energy does not decrease further than a value close
to $0.12$ and the specific heat decreases to a very small value. This
is the dynamical transition we have discussed in the previous
section. $T_G$ is well above the temperature $T_{RSB}$ and coincides
with the transition point derived for the marginality condition.

\begin{figure}
  \epsffile[60 206 565 690]{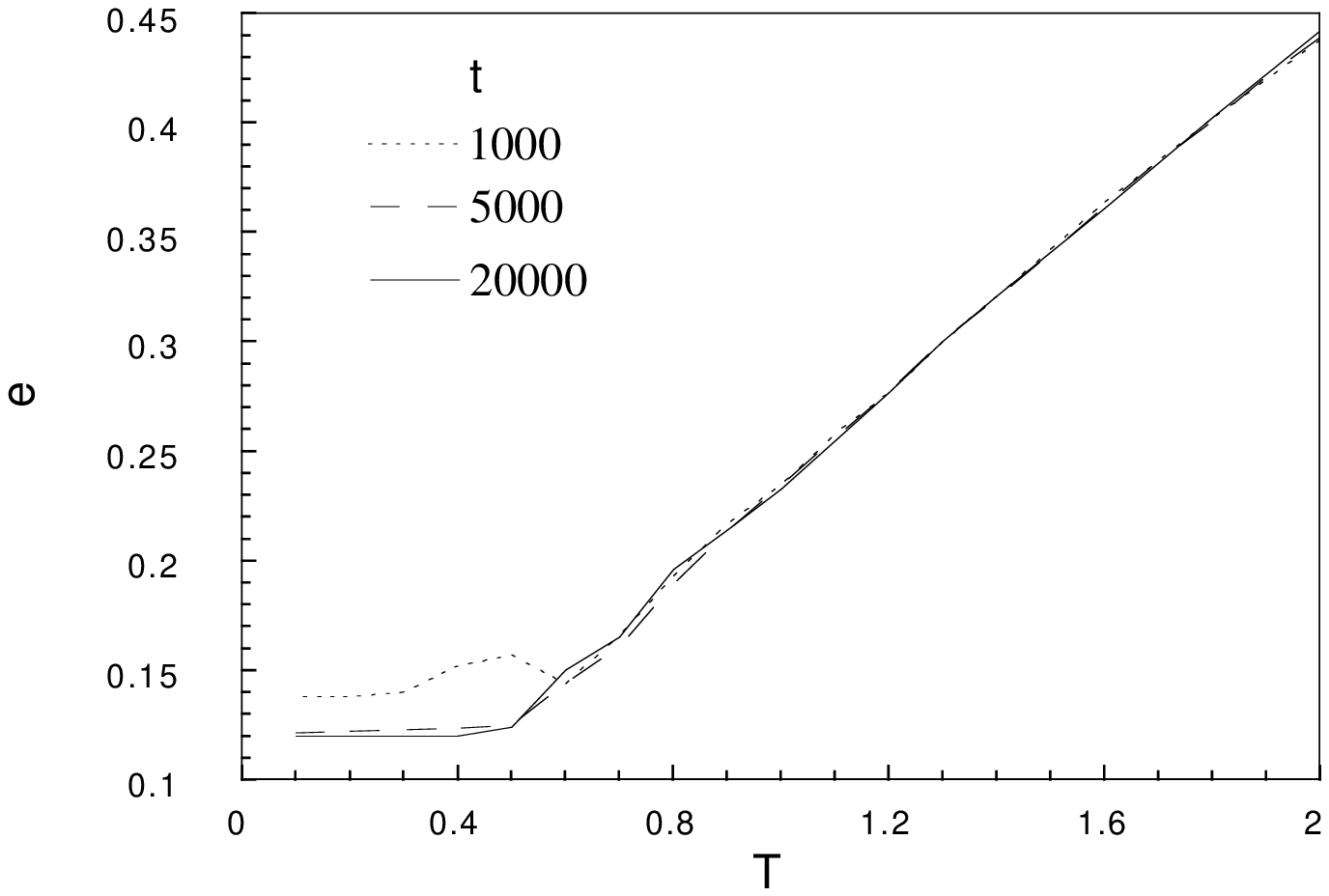}
  \caption[a]{\protect\label{F_ENETIME}
Energy of the model with quenched disorder versus $T$ for $N=100$.
Different curves correspond annealing schedules.
}
\end{figure}

The transition at $T_G$ is of a dynamical nature. The system does not
reach the lowest lying states (which have an energy close to $0.063$).
One could doubt if the freezing at $T_G\sim 0.5$ is a finite time
effect.  We show in fig. (\ref{F_ENETIME}) the internal energy of the
system as a function of $T$ (here $N=100$. We have used a value of $N$
not too small in order to to make the metastability visible). We plot
three different curves for different run length. In the run with
$t=1000$, for example, we sweep the lattice $1000$ times at each $T$
point during our annealing procedure (i.e. while systematically
decreasing $T$).

When the annealing time is too short for $T<T_G$ we get an energy that
is too high. But as soon as the scheduling becomes slow enough we see
that the energy thermalizes.  The dynamical freezing appears to be a
genuine behavior which survives in the limit of infinite times for
large volumes.  Let us note that for sizes less than $N\sim 50$ the
system is able to find the ground state in a reasonable time on our
simulation time scale, and we see it leaving the glassy phase.
The limits $N\to \infty$ and $t\to \infty$ seem not to commute.

\begin{figure}
  \epsffile[60 206 565 690]{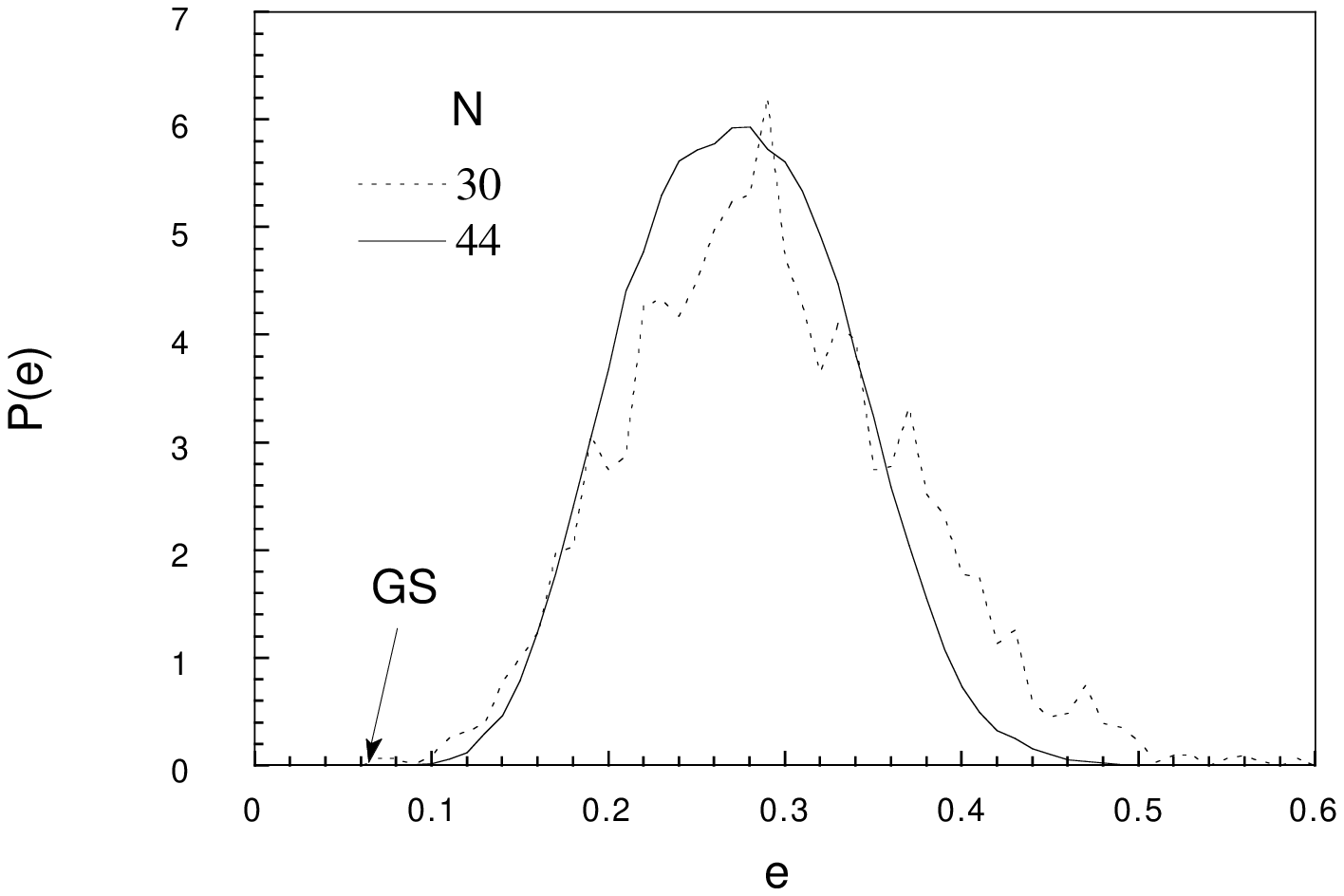}
  \caption[a]{\protect\label{F_ENEPROB}
Probability distribution of the energy of the metastable
states for size $N=30$ and $44$ for the model with quenched disorder. The
correct ground state energy is indicated with an arrow.
}
\end{figure}

Finally we show in figure (\ref{F_ENEPROB}) the distribution
probability for the energy of the metastable states at zero
temperature for quite small system size (where we are able to reach
the true ground state of the system).

For each lattice volume we have ran several millions of Monte Carlo
runs at zero temperature (we sweep sequentially the lattice and we
flip the local spin if so doing the internal energy decreases)
starting from different initial conditions and searching for
metastable states. We stop the search after we have found the lowest
energy state $100$ times.  We take that as good evidence for having
collected a fair sample of the low lying states.  In
fig. (\ref{F_ENEPROB}) we have also drawn an arrow locating the ground
state energy given by eq.(\ref{E_1SBFREE}) (which is close to
$0.063$). The agreement with our zero temperature results is
good. We also see that the distribution shape of the metastable states
is reminiscent of that found in case of the SK model \cite{BRAMOO}. We
will also see in the following that the energy distribution for the
deterministic model is similar to the one of the model with quenched
disorder (except for the existence of the very low lying ground state
we have written explicitly for certain values of $N$).

\section{Numerical Simulations of the Deterministic
Model\protect\label{S_NUD}}

We have studied the {\em cosine} model by using numerical simulations.
We will start by presenting results which describe the nature of the
ground state and illustrate the existence of a crystallization
transition for values of $N$ such that $(2N+1)$ is prime. Then we will
discuss the behavior of the internal energy and of the specific heat
during an annealing process.

\begin{figure}
  \epsfxsize=400pt\epsffile{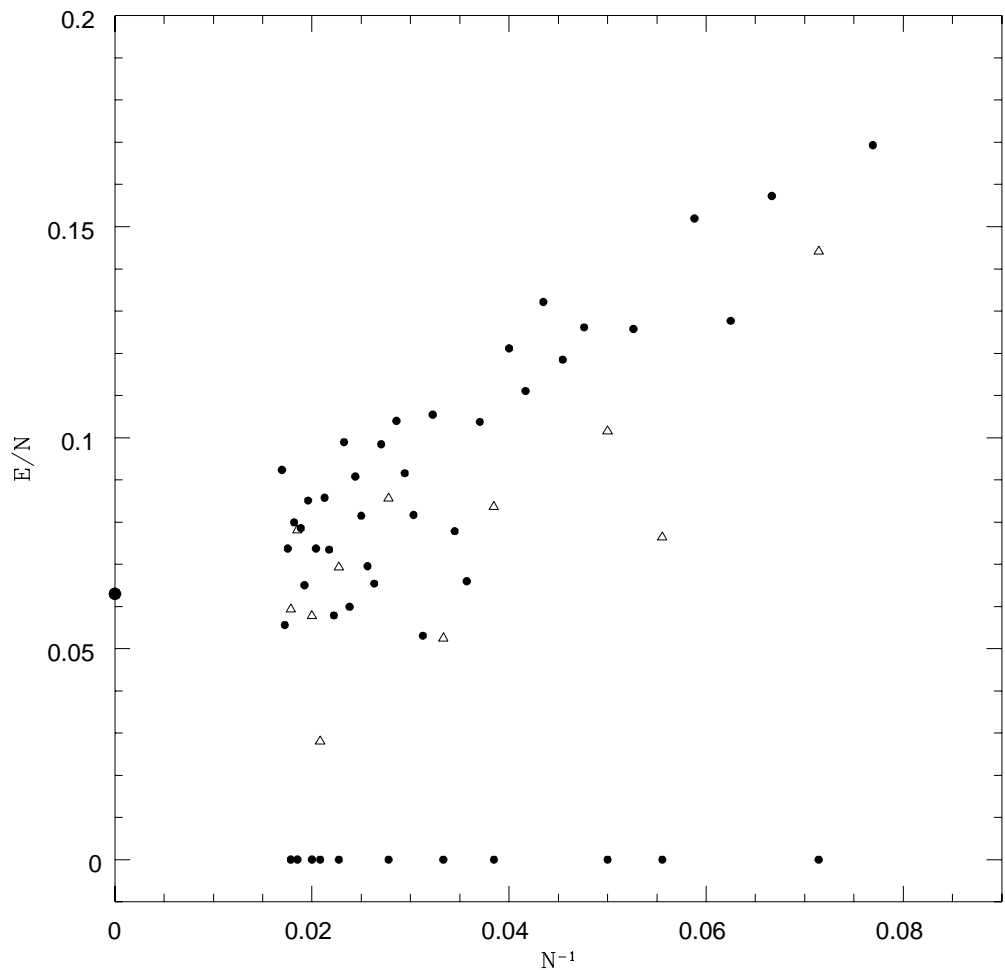}
  \caption[a]{\protect\label{F_EXAENE}
The ground state energy of the sine model divided by $N$ versus
$N^{-1}$ for small values of $N$ (at $N=\infty$ we plot the one
step replica broken analytic result we have obtained for ground state
of the model with quenched disorder). For $(2N+1)$ prime we also plot
the energy of first excited state with empty triangles.
}
\end{figure}

As we have discussed in section (\ref{S_GEN}) the Hamiltonian
(\ref{E_HQUA}) admits a zero energy ground state for values of $N$
such that $(2N+1)$ is prime.  We have found the ground state by exact
enumeration for small $N$ values (see \refA\ for a detailed discussion
of the technique).  For higher values of $N$ we have found the ground
state by looking for solutions of the naive mean field equations, as
we describe in the next section.  For finding the ground state this
method is slightly more efficient of the zero temperature Monte Carlo
introduced in the previous section.  In fig. (\ref{F_EXAENE}) we plot
the ground state energy divided by $N$ versus $N^{-1}$ for different
values of $N$ (at $N=\infty$ we plot the one step replica broken
analytic result we have obtained for the ground state of the model
with quenched disorder).  For $N$ such that $(2N+1)$ is prime we also
plot with a different symbol, the energy divided by $N$ of the first
excited state. The energy per spin is of order $0.1$.  The data of
fig. (\ref{F_EXAENE}) appear to be good evidence that for generic
values of $N$ the ground state energy tends to the value computed by
the replica approach (we suggest to the curious reader to compare
these results with the ones of \refA, since the difference is easy to
appreciate), and that the energy density does not vanish in the
thermodynamic limit. The excited states for $(2N+1)$ prime are a bit
lower than the ground state for generic $N$ values, but they do not
seem to have an atypical behavior.  In other words it would seem clear
that the pathology of the prime values $(2N+1)$ is confined to the
ground state. The spectrum of the higher energy states, including the
first excited state, does not depend on the cardinality of $(2N+1)$.

\begin{figure}
  \epsffile[60 206 565 690]{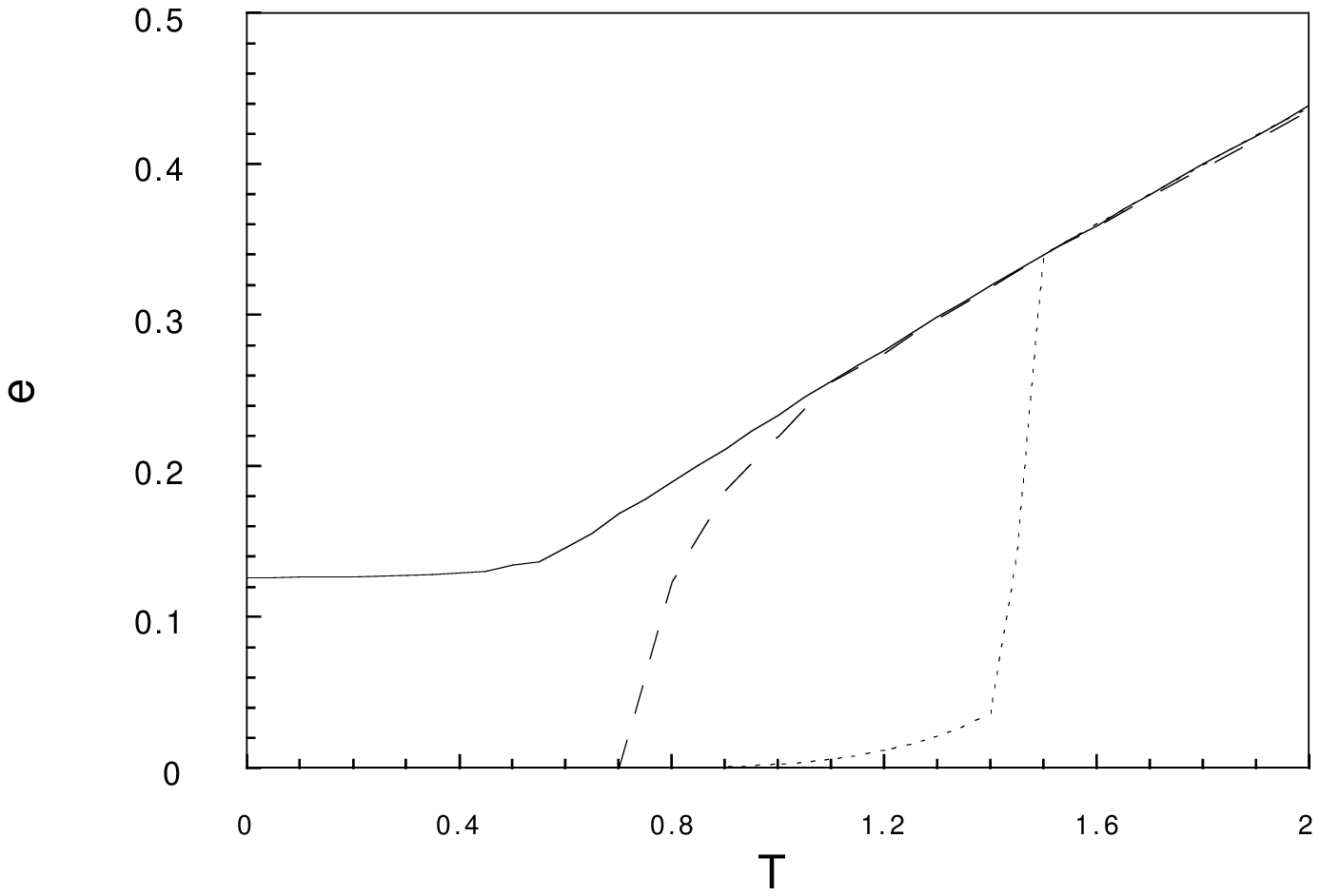}
  \caption[a]{\protect\label{F_HYSTER}
Energy of the cosine model versus $T$, for prime values of $(2N+1)$.
The solid line is for decreasing $T$ starting from a random
configuration (and is
the same for the two lattice sizes), while long dashes are starting
from a true ground state for
increasing $T$, $N=44$, and dots for increasing $T$, $N=806$.
}
\end{figure}

For prime values of $(2N+1)$ we find a crystallization first order
transition for $T_C\sim 0.7$. Knowing the exact form of the ground
state for such $N$ values has been a remarkable plus.  That allows us
to study the system both starting from high $T$ and cooling down to
low $T$ (in this case the system does never find the true ground
states, but gets trapped at the energy of the metastable phase) and
starting {\em from the ground state configuration}, slowly increasing
the temperature $T$. We are able in this way to observe a thermal
cycle we would not be able to detect in any other way. We show the
results (for $N=44$ and $N=806$, both such that $(2N+1)$ is prime) in
fig. (\ref{F_HYSTER}). The solid line is for decreasing $T$ (and is
the same for the two lattice sizes), while long dashes are for
increasing $T$, $N=44$, and dots for increasing $T$, $N=806$.

We notice that the area included between ascending and descending
curves increases with increasing $N$.  The crystallization transition
is of the first order, since the energy and the entropy are
discontinuous at $T_C$. The discontinuities $\Delta E$ and $\Delta S$
are such that $\Delta E=T_C\Delta S$. The free energy vanishes
approximately at $T_C$ (see figure \ref{F_RSFREE}) and remains very
close to zero below $T_C$ in the crystalline phase. In fact at low
temperatures the energy needed for a spin flip starting from the
ordered ground state is in the range $6-10$ so that the parameter for
a low temperature expansion of the free energy is of the order of
$\exp\{-\frac{6}{T}\}$. This means that the low temperature expansion
is well convergent and has a free energy which differs from zero by a
rather small amount in the whole region $T<T_C$.  The high temperature
free energy (given by (\ref{E_EQRS})) and the low temperature free
energy (which is equal to zero) intersect with an angle which is in
agreement with the first order nature of the crystallization
transition.

\begin{figure}
  \epsffile[60 206 565 690]{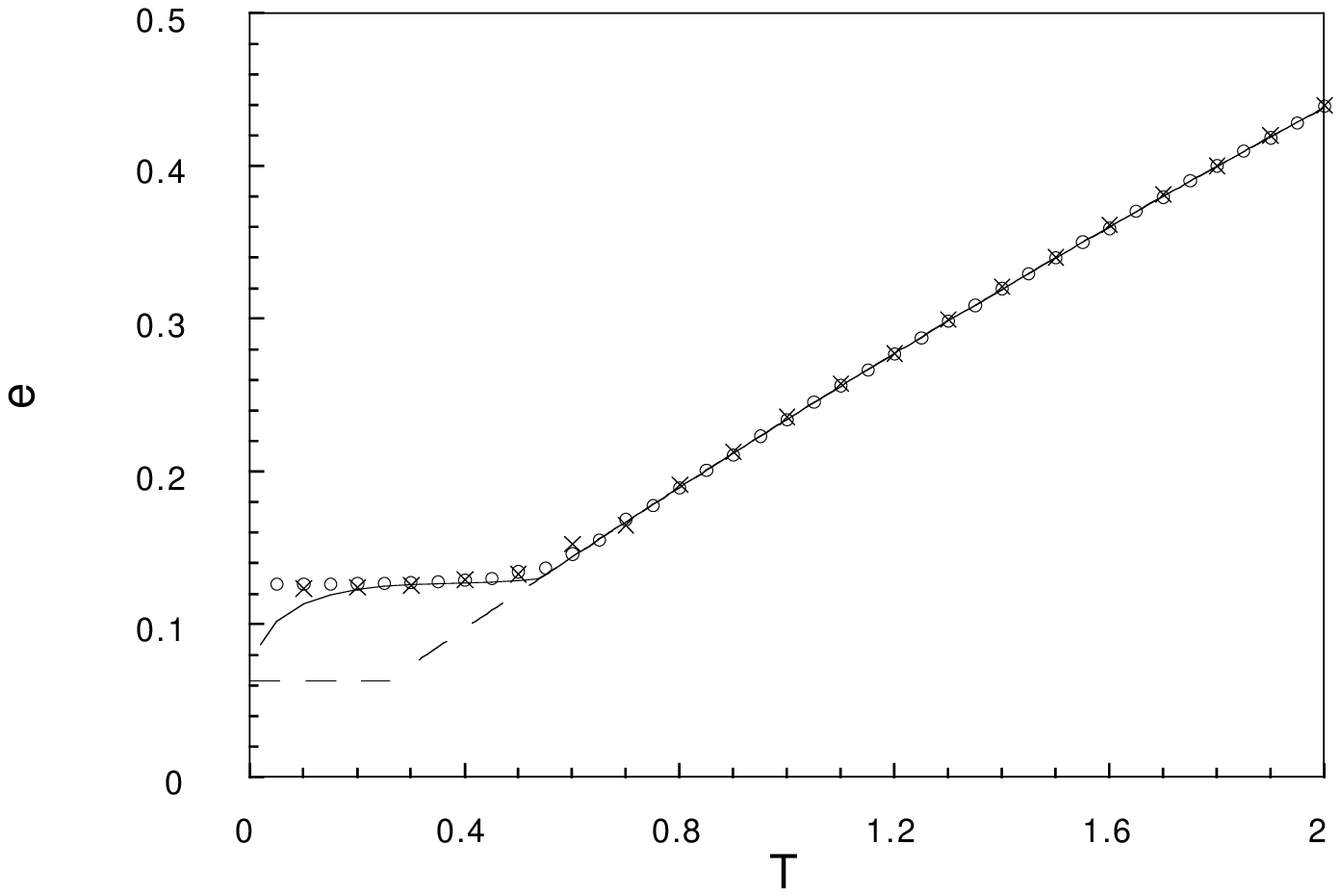}
  \caption[a]{\protect\label{F_ENDERA}
The internal energy as a function of $T$. Empty dots are from
numerical simulations for the cosine model, $N=806$. Crosses from numerical
simulations for the model with quenched disorder, $N=186$. The dashed line
is for the static one step replica symmetry broken solution, the
continuous line for the result obtained by imposing the marginality
condition.
}
\end{figure}

\begin{figure}
  \epsffile[60 206 565 690]{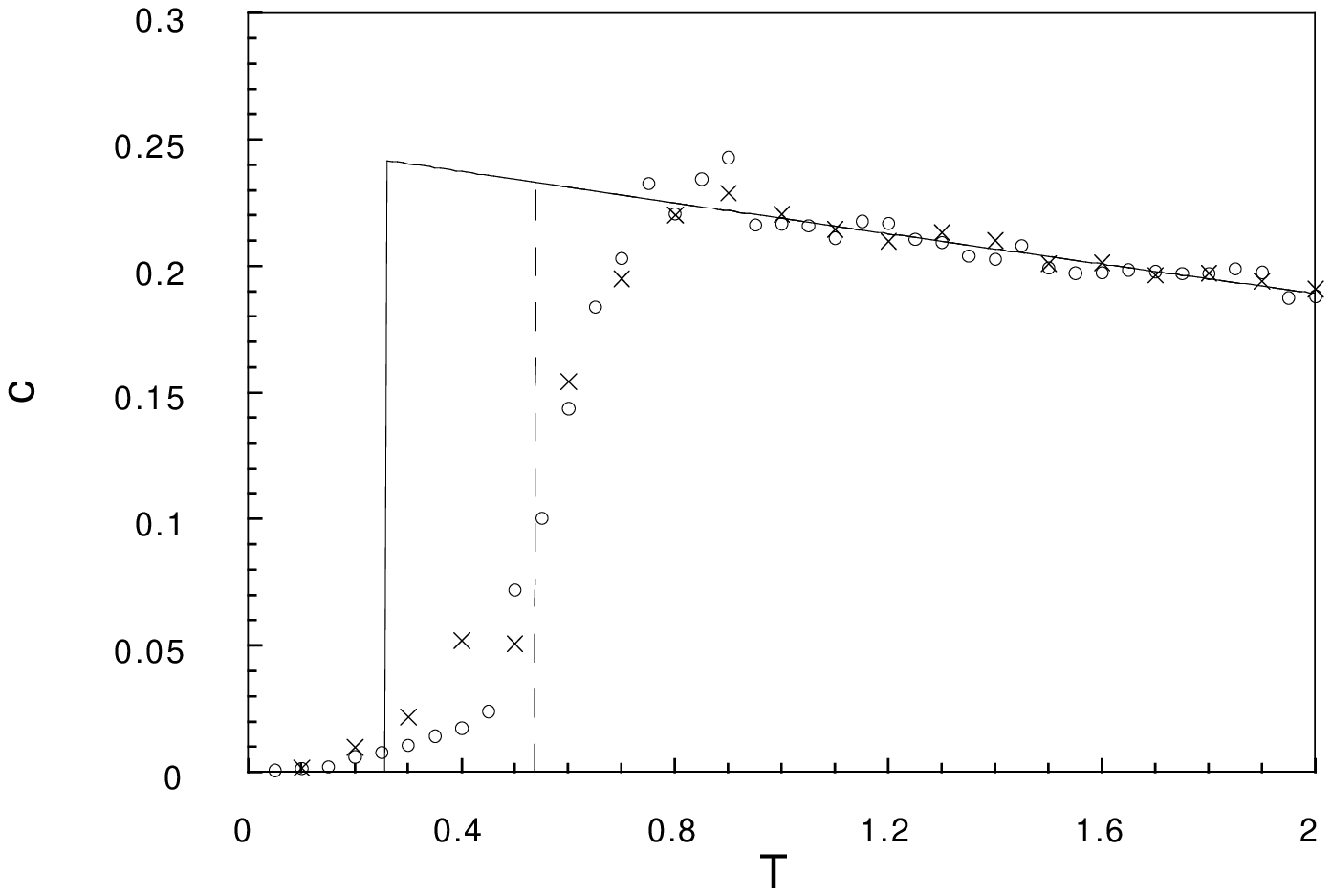}
  \caption[a]{\protect\label{F_CVDERA}
As in fig. (\ref{F_ENDERA}), but for the specific heat.
}
\end{figure}

Dynamically our system is able to undergo a crystallization transition
only for small values of $N$ which satisfy the cardinality condition. If
$(2N+1)$ is prime and $N$ is very large a local Monte Carlo annealing
dynamics is unable to bring the system in its true ground state.  The
system remains in a metastable phase exactly like it does in the model
with quenched disorder (where the zero energy ground state does not
exist).  In this regime the cardinality condition is irrelevant.  This
is illustrated by figures (\ref{F_ENDERA}) and (\ref{F_CVDERA}).  We
plot the energy and the specific heat versus $T$ for the cosine model
and for the model with quenched disorder (from numerical simulations),
for the one step broken solution and for the marginality condition
solution.

The model with quenched disorder has been conceived in order to
reproduce the high $T$ expansion of the deterministic model.  Below
the glass temperature $T_G$ there are no {\em a priori} reasons why the
two models should behave in a similar way.  The fact that the two
models coincide also in the metastable phase is clear from the results
we show in figures (\ref{F_ENDERA}), (\ref{F_CVDERA}), and comes as a
very nice surprise.  One of the reasons for such a behavior is the
fact that the metastable states in the two models have a very similar
distribution, as we will show better in next section.

Also in case of the deterministic model, the metastable phase can be
described using the marginality condition eq.(\ref{E_MARG}) of the
section $6$.

Figure (\ref{F_ENDERA}) shows that the solution where the marginality
condition has been imposed describes very well the numerical results
down to $T\sim 0.1$. Below that temperature the energy of the analytic
solution departs from the numerical results reaching the static value
$\sim 0.063$ at $T=0$. This behavior is related to the fact that the
breakpoint parameter $m$ (as determined by imposing the marginality
condition) is not proportional to $T$ for low values of $T$. This fact
will be discuss in better detail in the appendix and confirms the fact
that the static replica equations are useful to predict the existence
of the glassy transition at $T_G$ but possibly not the full low $T$
region.

The next section is devoted to describe the structure of the
metastable states for the deterministic model at zero temperature by
analyzing the numerical solution of the naive TAP equations. But for
the existence of a crystalline state in case of prime $(2N+1)$ prime
the shape of the distribution of the metastable states will be shown
to be similar to the one found in case of the random model.

\section{Mean Field Equations for the
Deterministic Model\protect\label{S_MF}}

The naive mean field equations for the sine model can be defined
through the iterative relation

\be
  m_x = \tanh(\beta\sum_{y\ne x}S_x(m_y)  )\ ,
\ee

\noindent
where the function $S_x$ has been defined in (\ref{E_S}). Obviously we
could have defined the analogous equations by using the $C$ function
defined in eq. (\ref{E_C}).

We are interested in the low temperature limit of the model.  We can
thus avoid to consider the complete TAP equations, where the reaction
field is included, which are far more difficult to deal with. In the
low temperature limit we can solve the even simpler equations

\be
  m_x = \mbox{\rm sign} (h_x)\ ,
\ee

\noindent
where $h_x$ is the local field acting on the spin $x$.

\begin{figure}
  \epsfxsize=400pt\epsffile{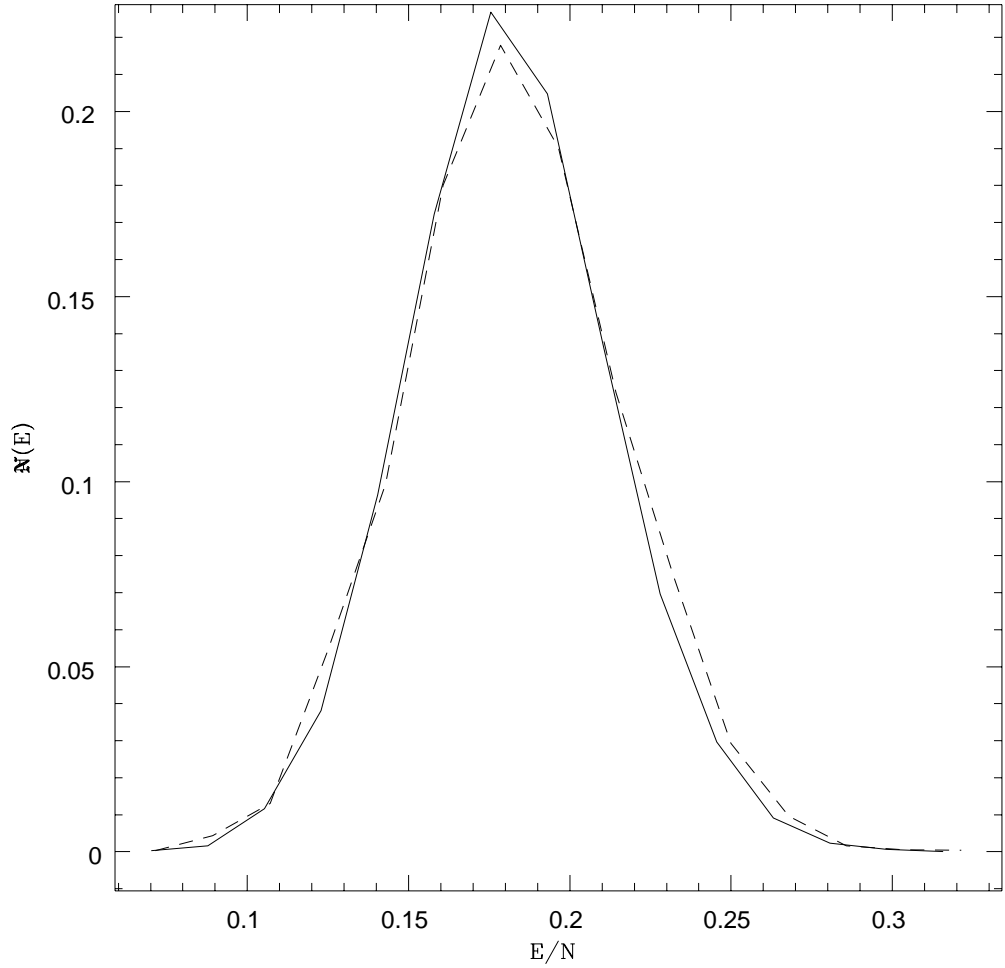}
  \caption[a]{\protect\label{F_MF}
  The number of solutions of the $T=0$ mean field equations of a given
  energy as function of the energy for $N=56$, where $(2N+1)$ is prime
(dashed line) and for $N=57$, where $(2N+1)$ is not prime (continuous line).
}
\end{figure}

We find the $T=0$ solution of these equation by cooling the solution
found at $T>0$.  In figure (\ref{F_MF}) we show
the number of solutions of a given energy as function of the energy
respectively for a typical prime (dashed line) and non prime value
(continuous line)of $p=(2N+1)$.

Analogously to \refA\ we stop our ground state search after finding
$5$ times the states with the lowest energies. That makes us confident
we have sampled the low energy states with good accuracy.
We have studied systems with $N$ up to $64$. For prime values of
$(2N+1)$, where we know the exact ground state, this method has always
found the correct ground state energy (i.e. zero).

\section{Conclusions\protect\label{S_CON}}

Building upon the idea introduced in our former paper \refA\ we have
introduced here a class of deterministic spin models which do not
contain disorder, but whose low $T$ behavior is dictated by
self-induced frustration. They are potentially relevant to the
description of the glass state.  Using number theory we have been able
to exhibit a zero energy ground state for given values of the volume
$N$ (such that $(2N+1)$ is prime).

We have proceeded by writing a model with quenched random disorder,
based on orthogonal interaction matrices, which reproduces the high
temperature expansion of the deterministic models. By using replica
theory and well known results of integration on Lie groups we have
been able to solve the model with quenched disorder.
The model with quenched disorder has a replica symmetry breaking
transition at a quite low temperature. The phase transition
is discontinuous like in the random energy model.

We have also studied the low $T$ phase.  Even if the random model does
not coincide with the deterministic model for all values of $N$ down
to $T=0$ (since we know that for prime values of $(2N+1)$ the
deterministic model admits a ground state based on Legendre sequences
which we cannot find in the random approach) we have found that in all
the metastable phase th two class of models coincide. We have also
found, remarkably, that for generic values of $N$ even the ground
states of the models seem to coincide (as from figure
(\ref{F_EXAENE})).

We have shown that for the values of $N$ which satisfy the cardinality
condition the deterministic model undergoes a crystallization
transition. This transition is of the first order from the
thermodynamical point if view, since the energy and the
entropy jump discontinuously. Even if we cannot be sure of this fact,
our exact solutions of small systems give a precise hint favoring the
absence of a zero energy ground state for generic $N$ values.

We have shown that the structure of metastable states of the two
classes of models has much in common (at this effect the cardinality
of $N$ is irrelevant).  For the model with quenched disorder we have
performed Monte Carlo runs at zero temperature searching for locally
stable states. In the deterministic case we have solved the naive TAP
equations.  The similarity of the shapes of the distribution of
metastable states suggests that the dynamical behavior of the two
models must be very similar. The two figures
(\ref{F_ENDERA}), (\ref{F_CVDERA}) are quite decisive in this respect.
The two models behave very similarly, they both display a singularity
at a temperature $T_G$ where the system freezes and thermodynamic
fluctuations (related for example to the specific heat and to the
magnetic susceptibility) vanish.
We have also shown that for reasons that are quite unclear to us
the marginality condition gives a good estimate of the low $T$
behavior.

These results strengthen the idea that the off-equilibrium dynamics
for the deterministic model should be very similar to the one of the
model with quenched disorder.  We would expect, for example, that the
deterministic model could display aging effects like those which
affect the random model and many models based on quenched disorder
\cite{AGING}. We have measured the usual time-time correlation
function between the spin configuration at the waiting time $t_w$ and
the spin configuration at a later time $t_w+t$.  We have observed that
below $T_G$ the shape of the correlation function depends on the
previous history, i.e. on $t_w$. These results are much similar to
those found also in related deterministic models like the low
autocorrelation binary sequences \cite{MIGRIT}.  It seems that also
deterministic models display non-equilibrium effects very similar to
those of spin glasses with randomness.

We hope that the results of this paper can be relevant to a large variety of
different problems in condensed matter physics, where it is natural to
study systems with a complex free energy landscape in which
quenched disorder is not present as a given, preassigned condition.

\section*{Acknowledgements}

We thank Andrea Crisanti, Leticia Cugliandolo and Jorge Kurchan for
discussions.  G.~P. thanks Bernard Derrida for stressing to
him the interest of computing the average minimal distance of an Ising
spin configuration from a random hyperplane.

\vfill

\newpage

\section*{Appendix\protect\label{S_APP}}

In this appendix we present some technical details about how we
applied the marginality condition in our computation at one step of
replica symmetry breaking. Our starting point is the expression for
the free energy

\be
  \protect\label{main}
  A[Q,\La] =
  -\frac{1}{2}\  \mbox{\rm Tr} G(2\beta Q)+ \mbox{\rm Tr}  (\La Q) -F(\La)\ ,
\ee

\noindent with $G(Q)$ given by

\be
G(2\beta Q)=\sum_{k\ge 1}\frac{(2\beta)^{2k}}{2k}\psi_{2k}\,\mbox{\rm Tr}
  \mbox{\rm Tr} Q^{2k}=\\
\sum_{k\ge 1} c_{2k}\,\mbox{\rm Tr} Q^{2k}\ ,
\ee

\noindent
where the $\psi_{2k}$ are the Taylor coefficients of the series expansion
of the function $\psi (z)$:

\be
\psi (z)=1+\sum_{k\ge 1}\psi_{2k}\,z^{2k}\ .
\ee

\noindent
In the most general case
the stability condition implies that the Hessian
matrix of the second derivatives of $A[Q,\La]$ in the space of matrices
$\lbrace Q,\La\rbrace$ around the equilibrium solution is negative
definite (the integration path in
$\La$ space runs on is the imaginary axis, and the stability condition
has the opposite sign than in the usual case).
To construct the Hessian we compute the second derivatives of
(\ref{main}).
This gives a four blocks matrix with the derivatives
$\partial_{QQ} A$ , $\partial_{\La\La} A$ , $\partial_{Q\La} A$ and the
identical symmetric block $\partial_{\La Q} A$.  The sub-block $G\equiv
\partial_{QQ} A$ is given by

\be
G_{(ab)(cd)}=\frac{\partial^2 A}{\partial Q_{ab}\partial Q_{cd}}=
\sum_{k\ge 1} 4k c_{2k} \frac{\partial (Q^{2k-1})_{ab}}{\partial
Q_{cd}}\ .
\ee

\noindent
The matrix $G$ has three different types of elements, depending on if
the replica indices $(ab)$ and $(cd)$ do coincide, have one element in
common or are completely different. For these three different cases we
have

\bea
\frac{\partial (Q^{2k-1})_{ab}}{\partial Q_{cd}}=\sum_{p=0}^{2k-2}
\Bigl ( (Q^p)_{ac}(Q^{2k-2-p})_{db}\,+\,(Q^p)_{ad}(Q^{2k-2-p})_{cb} \Bigr )\\
\frac{\partial (Q^{2k-1})_{ab}}{\partial Q_{ac}}=\sum_{p=0}^{2k-2}
\Bigl ( (Q^p)_{aa}(Q^{2k-2-p})_{cb}\,+\,(Q^p)_{ac}(Q^{2k-2-p})_{ab}
\Bigr )\\
\frac{\partial (Q^{2k-1})_{ab}}{\partial Q_{ab}}=\sum_{p=0}^{2k-2}
\Bigl ( (Q^p)_{aa}(Q^{2k-2-p})_{bb}\,+\,(Q^p)_{ab}(Q^{2k-2-p})_{ab} \Bigr )
\nonumber \ .
\eea

\noindent
The other sub-blocks $I$ and $M$ are

\bea
I_{(ab)(cd)}&=&\frac{\partial^2 A}{\partial \La_{ab}\partial Q_{cd}}=
\delta_{(ab)(cd)}\\
M_{(ab)(cd)}&=&\frac{\partial^2 A}{\partial \La_{ab}\partial \La_{cd}}=
\langle \sigma_a\sigma_b\rangle_F\langle\sigma_c\sigma_d\rangle_F-
\langle \sigma_a\sigma_b\sigma_c\sigma_d\rangle_F\ .
\eea

\noindent
The mean value $\langle...\rangle_F$ in the last equations is taken
over the action (\ref{E_FLA}).  $M$ is
the usual Hessian which determines the stability of the SK model.

Now it is easy to see that for each eigenvalue of the sub-block matrices
$G$ and $M$, (for instance $g$ and $\mu$ respectively) the stability
condition is determined by

\be
  g\mu -1 \le 0\ ,
  \label{stab}
\ee

\noindent
the marginal condition being the equality. We have now to compute all
the eigenvalues of the matrices $G$ and $M$ and to search for the ones
which maximize the product $g \nu$. For the $p$-spin model (and also
the SK model) this condition is relatively easy to determine because
there is a unique eigenvalue $g$ for $G$ (in that case the matrix $G$
is $g$ times the identity matrix) and the maximum eigenvalue of $M$ is
found in the replicon sector when all replicas belong to the same
block (once replica symmetry is broken).

In the present case even though the maximum
value of $M$ is the usual one \cite{Go} $G$ has more than one
eigenvalue.  We have searched for all of them in case of one step of
replica symmetry breaking. We have evaluated the derivatives
for the matrices $Q$ and $\La$ broken according
to the scheme of (\ref{break}). The general expression for the
eigenvalues at one step of replica symmetry breaking has been
given in \cite{Bru}. There are two longitudinal eigenvalues, four
anomalous eigenvalues and four replicons which finally reduce to only
five different eigenvalues (this is because we set $Q_{(ab)}=0$ if the
indices $(a,b)$ do not belong to the same sub-block of size $m$). These
are given by

\bea
g_1&=&\frac{16\beta^2}{m}(G''(4\beta (1-q))+(m-1)G''(4\beta(mq+1-q)))\\
g_2&=&16\beta^2 G''(4\beta (1-q+mq))\\
g_3&=&\frac{32\beta^2}{m}G''(4\beta (1-q))+\frac{4\beta (m-2)}{qm^2}
(G'(4\beta (1-q+mq))-G'(4\beta(1-q)))\\
g_4&=&\frac{4\beta}{qm}
(G'(4\beta (1-q+mq))-G'(4\beta(1-q)))\\
g_5&=&16\beta^2 G''(4\beta (1-q))\ .
\eea

\noindent
$g_5$ is the replicon, where all the replica indices belong to the
same sub-block.  Taking for the matrix $M$ the replicon eigenvalue
corresponding to the four replica indices all belonging to the same
sub-block we find

\be
\mu=\langle\cosh^{-4}(\sqrt{2\lambda}x)\rangle\ ,
\ee

\noindent
where the expectation value is defined by

\be
\langle A(x)\rangle =\frac{\int dx \frac{e^{-x^2}}{\sqrt{2\pi}}
\cosh^m(\sqrt{2\la}x)A(x)} {\int dx \frac{e^{-x^2}}{\sqrt{2\pi}}
\cosh^m(\sqrt{2\la}x)}\ .
\ee

\noindent
Inserting in (\ref{stab}) this value of $\mu$ we have searched among
the $5$ values of $g$ the one which gives the maximum free energy when
the stability is marginal (i.e. when (\ref{stab}) is saturated). We
have found that the eigenvalue $g_5$ is the one which gives the
maximal free energy . This leads us to the marginality condition
(\ref{E_MARG}).  We have searched for a solution of the marginality
condition in which $m$ behaves linearly with $T$ for low temperatures,
but we have not been able to find it. It is plausible that such well
behaved solution does not exist and that to improve our solution one
would need to break the replica symmetry with a larger number of
steps.

\vfill
\newpage

\end{document}